\documentclass[sigconf,nonacm]{acmart}

\usepackage[normalem]{ulem}
\usepackage{caption,xspace}
\usepackage{pifont}
\usepackage{textcomp,balance}
\usepackage{bm}
\usepackage{amsfonts}
\usepackage{url}
\usepackage{multirow}
\usepackage{booktabs}
\usepackage{subcaption} 
\usepackage{booktabs}
\usepackage{tabularx}
\usepackage{numprint}
\usepackage{ragged2e}
\usepackage{setspace}
\usepackage{framed}
\usepackage{wrapfig}
\usepackage{tcolorbox}
\usepackage{makecell}

\usepackage[linesnumbered,ruled,vlined]{algorithm2e}

\usepackage{amsmath,amssymb}
\usepackage{algorithmic}
\usepackage{graphicx}
\usepackage{xcolor}
\usepackage{tikz}
\usepackage{listings,balance,lstautogobble}
\usepackage{cleveref}
\usepackage{pifont}
\usepackage{soul}
\usepackage{enumitem}
\usepackage{kantlipsum}
\usepackage{setspace}
\usepackage{colortbl}
\usepackage{adjustbox}

\Crefname{figure}{Fig.}{Figs.}
\Crefname{equation}{Eq.}{Eq.}
\Crefname{section}{Sec.}{Sec.}
\newcommand{\system}{QCompass\xspace}
\newcommand{\bench}{QuOptBench\xspace}

\definecolor{darkgreen}{RGB}{0,120,0}

\definecolor{lightred}{rgb}{1.0,0.85,0.85}
\definecolor{lightblue}{rgb}{0.85,0.85,1.0}

\definecolor{costHigh}{rgb}{0.8,0.85,0.8}
\definecolor{costMid}{rgb}{0.9,0.95,0.9}
\definecolor{costLow}{rgb}{0.95,0.99,0.95}

\definecolor{gainHigh}{RGB}{220,80,80}   
\definecolor{gainMid}{RGB}{240,150,150}  
\definecolor{gainLow}{RGB}{250,220,220}  

\makeatletter
\g@addto@macro\normalsize{%
  \setlength{\abovedisplayskip}{2pt plus 1pt minus 1pt}%
  \setlength{\belowdisplayskip}{2pt plus 1pt minus 1pt}%
  \setlength{\abovedisplayshortskip}{2pt plus 1pt minus 1pt}%
  \setlength{\belowdisplayshortskip}{2pt plus 1pt minus 1pt}%
}
\makeatother
\normalsize

\AtBeginDocument{%
  }

\begin{document}

\title{Turning Quantum Compiler Configuration Selection Upside Down: A Top-Down Paradigm }

\author{Ming Zhong}
\email{mzhong25@cse.cuhk.edu.hk}
\orcid{0009-0002-7814-7523}
\affiliation{%
  \institution{The Chinese University of Hong Kong}
  \country{Hong Kong SAR, China}
}

\author{Xiangyu Ren}
\orcid{0009-0006-8914-2188}
\email{xiyu.ren@outlook.com}
\affiliation{%
  \institution{The Chinese University of Hong Kong}
  \country{Hong Kong SAR, China}
}

\author{Kangyu Zheng}
\orcid{0009-0008-1165-0814}
\email{kyzheng@cse.cuhk.edu.hk}
\affiliation{%
  \institution{The Chinese University of Hong Kong}
  \country{Hong Kong SAR, China}
}

\author{Zhi Zeng}
\orcid{0009-0002-3471-6405}
\email{zzeng26@cse.cuhk.edu.hk}
\affiliation{%
  \institution{The Chinese University of Hong Kong}
  \country{Hong Kong SAR, China}
}

\author{Shenggen Zheng}
\orcid{0000-0001-7325-2506}
\email{zhengshenggen@gmail.com}
\affiliation{%
  \institution{Quantum Science Center of Guangdong-Hong Kong-Macao Greater Bay Area}
  \country{Guangdong Province, China}
}

\author{Shaohua Li}
\orcid{0000-0001-7556-3615}
\email{shaohuali@cse.cuhk.edu.hk}
\affiliation{%
  \institution{The Chinese University of Hong Kong}
  \country{Hong Kong SAR, China}
}

\author{Zhiding Liang}
\authornote{Corresponding Author.}
\orcid{0000-0002-7568-0165}
\email{zliang@cse.cuhk.edu.hk}
\affiliation{%
\institution{The Chinese University of Hong Kong; State Key Laboratory of Quantum Information Technologies and Materials (CUHK)}
\country{Hong Kong SAR, China}
}

\begin{abstract}
Selecting an effective quantum compiler configuration requires coordinated decisions over both hardware targets and compiler settings, with the ultimate goal of improving circuit execution fidelity. Existing state-of-the-art approaches such as MQT Predictor typically decompose this problem into two sequential stages: a device is chosen first, after which compiler options are tuned on that device through iterative reinforcement-learning-based search. This \emph{bottom-up} strategy can be fundamentally limited by its initial device decision, since a poorly matched backend may constrain the best fidelity achievable regardless of how effectively later compiler options are optimized. In contrast to classical compiler tuning, where the search space is often prohibitively large, the configuration space of practical quantum compilation remains manageable enough to support exhaustive offline evaluation. This property opens up a different possibility: rather than repeatedly searching from candidate configurations, one can derive reusable configuration knowledge from offline oracle results and apply it directly to unseen circuits.

Based on this idea, in this paper, we introduce a \emph{top-down} \linebreak configuration-selection paradigm and build \bench, an exhaustively characterized benchmark containing empirical oracle configurations for 268 quantum circuits. We first study whether LLMs can directly infer suitable configurations and evaluate ten representative models. The results reveal that vanilla LLMs often favor unnecessarily complex compiler settings and therefore remain far from the oracle. We then employ Claude-Opus-5 to examine oracle decisions from 68 sampled circuits and extract 13 transferable rules governing device and compiler-option selection. Using these rules together with lightweight predictors trained on the complete enumerated configuration data of the sampled circuits, we design \system, a compact rule-guided ML framework that performs fast configuration selection without iterative online search. Across our evaluation, the distilled rules consistently strengthen LLM-based selection, while \system provides the strongest overall balance between solution quality and selection latency among the compared methods. These results indicate that transferring knowledge from offline oracle characterization offers a promising alternative to conventional bottom-up quantum compiler tuning.

\end{abstract}

\maketitle

\section{Introduction}
\label{sec:intro}

Quantum computing has rapidly advanced with the development of both quantum hardware and software~\cite{PhysRevLett_qa, quantum_chem,osti_1623945,liang2024combining}. Unlike compilation for classical programs, quantum compilation must account for noisy and error-prone quantum hardware. Therefore, the quality of a compiled circuit on a target device is commonly measured by execution \emph{fidelity}~\cite{Jozsa01121994}, which reflects the combined effects of compilation and hardware characteristics, such as topology and gate errors.

\begin{figure*}[t]
    \centering
    \begin{subfigure}[t]{0.54\linewidth}
        \centering
        \includegraphics[width=\linewidth]{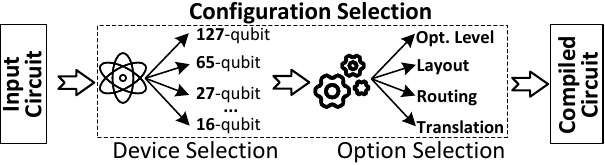}
        \caption{Two-stage quantum compiler configuration selection.}
        \label{fig:selection}
    \end{subfigure}
    \hfill
    \begin{subfigure}[t]{0.45\linewidth}
        \centering
        \includegraphics[width=\linewidth]{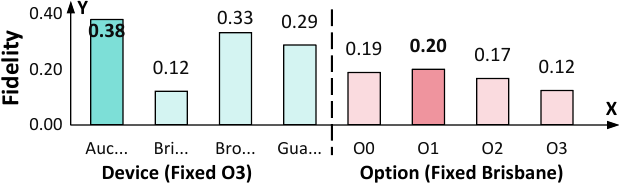}
        \caption{Fidelity variation for the \texttt{ising\_n10} circuit.}
        \label{fig:motivation_o3}
    \end{subfigure}
    \caption{Workflow and motivation for quantum compiler configuration selection.}
    \Description{An overview of device and compiler option selection in quantum compilation and an example showing fidelity differences caused by different devices and optimization levels.}
    \label{fig:mot_tea}
    \vspace{-.ex}
\end{figure*}

As illustrated in \Cref{fig:selection}, quantum compilation involves two closely coupled decisions. The first is \textbf{\emph{device selection}}. Since different quantum devices have distinct gate error rates and connectivity topology, selecting a suitable device for each circuit is critical to maximizing execution fidelity, analogous to routing requests to appropriate LLMs~\cite{feng2026llmrouter}. The second is \textbf{\emph{option selection}}, which determines how the circuit is compiled for the selected device through stages including layout, routing, translation, etc. Importantly, first-stage device decision defines the hardware constraints and error characteristics for all subsequent compilation stages.

Taking the widely used IBM Qiskit compiler~\cite{Web:qiskit} as an example, users can configure optimization levels from $O_0$ to $O_3$, as well as methods including layout, routing, etc. Although $O_3$ applies the most complex optimization flow, neither $O_3$ nor any method is universally optimal across circuits.

\Cref{fig:motivation_o3} illustrates the impact of both device and option selection. Under a fixed $O_3$ configuration, the \texttt{ising\_n10} circuit achieves substantially different execution fidelities (measured by Hellinger fidelity~\cite{fidelity}) across devices; on a fixed device, varying the optimization level from $O_0$ to $O_3$ also leads to clear differences, with $O_1$ achieving about 0.20 fidelity compared with only 0.12 for $O_3$. This shows that neither a fixed device nor the $O_3$ option is consistently optimal.

This observation is further supported by our later evaluation, where we exhaustively enumerate the considered devices and compiler options for 268 circuits. Under our compilation settings, 213 cases (79.5\%) have at least one capacity-feasible device whose best achievable fidelity (measured by estimated fidelity~\cite{mqtpredictor,DAC23}) across all option combinations is still lower than the default $O_3$ configuration fidelity on another feasible device. This result shows that an unfavorable device choice can impose a fidelity ceiling that subsequent option selection cannot overcome.

Therefore, we formulate the joint selection of suitable devices and compiler options for quantum circuits as quantum compiler \textbf{configuration selection}. In classical compilers, a similar problem is \emph{compiler auto-tuning}, which has been addressed using heuristic search~\cite{boca,cobyan,PDCAT,gene1,gene2,gene3}, reinforcement learning (RL)~\cite{rl1,rl2,rl3,rl4}, and more recently LLMs~\cite{pan2026eccoevidencedrivencausalreasoning,qiu2026passbypassoptimizationintentdrivenir,meta_llm_compiler}. Heuristic and RL-based methods typically start from candidate configurations and iteratively improve them using predefined rules or performance feedback, even though a trained RL agent can perform direct inference. We refer to this as a \textbf{\emph{bottom-up}} search paradigm, whose selection knowledge is acquired by iteratively evaluating candidate configurations and learning from feedback. In contrast, LLMs can directly reason about suitable configurations, resembling a direct selection process.

{
\[
\underbrace{
\text{Candidate Configurations}
\rightarrow
\text{Evaluation}
\rightarrow
\text{Better Configuration}
}_{\text{\textbf{Bottom-Up Search}}}
\]
}

Additionally, classical compiler auto-tuning typically assumes a fixed target platform and may involve extremely large flag or pass spaces, e.g., up to 126 flags ($2^{126}$ search space) in a recent GCC auto-tuning study~\cite{qiao2026warmtunerprogramspecificwarmstarts}. Quantum compiler configuration differs in two important aspects. First, the target device is itself a selection dimension. The state-of-the-art (SOTA) method~\cite{mqtpredictor} adopts a two-stage workflow, using an ML model for device selection and an RL model for compiler option tuning.  While this modular design facilitates new-device integration, a suboptimal  device choice can impose a fidelity ceiling that subsequent option tuning cannot overcome.

Second, quantum compiler configuration has a bounded discrete search space in widely used compilers such as Qiskit~\cite{Web:qiskit}. In our setting, each circuit has only 960 device-option configurations, making exhaustive offline enumeration feasible for obtaining empirical oracles. This enables an oracle-guided \textbf{\emph{top-down selection}} paradigm, where selection rules extracted from these oracle configurations directly guide configuration selection without bottom-up search.

{
\[
\underbrace{
\text{Oracle Configurations}
\rightarrow
\text{Rules}
\rightarrow
\text{Direct Selection}
}_{\text{\textbf{Top-Down Selection}}}
\]
}

In this paper, we systematically investigate the proposed \emph{top-down} selection paradigm using the Qiskit compiler. We first construct \bench, the first \emph{oracle-guided} benchmark for Qiskit compiler configuration selection, covering 268 circuits with empirical oracle configurations. We then evaluate ten LLMs as a natural realization of direct selection. Results show that vanilla LLMs remain substantially below the oracle, largely due to their preference for overly aggressive configurations, such as higher optimization levels and more complex layout and routing methods.

Motivated by this finding, we instead use a strong reasoning model, Claude-Opus-5~\cite{Web:claude}, to analyze oracle decisions on 68 carefully categorized and sampled circuits and distill 13 transferable device- and option-selection rules. Based on these rules, we develop \system, a lightweight rule-guided ML framework for selecting configurations for new circuits. Experimental results show that the distilled rules consistently improve LLM-based selection, while \system achieves higher fidelity with a favorable trade-off between configuration quality and inference overhead.

The main contributions of this work are summarized as follows:

\begin{itemize}[leftmargin=*]
    \item We construct \bench, the first exhaustive benchmark for the Qiskit quantum compiler configuration selection containing 268 circuits with empirical oracle configurations.

    \item We investigate ten LLMs as a natural realization of direct quantum compiler configuration selection. Our empirical study shows that general-purpose LLM reasoning alone cannot reliably reproduce oracle-quality configurations.

    \item We develop an oracle-guided rule-distillation workflow that contrasts $O_3$ configurations with exhaustive-search oracles, explains their differences using circuit and device characteristics, and consolidates recurring patterns into transferable device- and option-selection rules.

    \item We design \system, a lightweight rule-guided learning and inference approach that applies the distilled rules to new circuits, shifting quantum compiler optimization from repeated bottom-up search to efficient top-down selection.

    \item We conduct comprehensive evaluations of \system, including comparisons with bottom-up search methods, nearest-neighbor retrieval, and direct LLM-based selection, as well as generalization to additional circuits and devices.
\end{itemize}

\section{\bench: An Oracle Benchmark for Quantum Compiler Configuration Selection}\label{sec:bench}

\subsection{Task Description}\label{sec:tasks}

The task of \bench is to jointly select a device and compiler-option combination for each circuit. Following prior work~\cite{DAC23,mqtpredictor}, the selection is based on circuit- and device-level features. Circuit features cover seven metrics: qubit count, depth, program communication, critical depth, entanglement ratio, parallelism, and liveness. Device features characterize hardware capacity, noise, and connectivity, including qubit count, two-qubit and readout errors, $T_1$/$T_2$, average degree, and edge density. Together, these features provide the information needed for configuration selection.

\subsection{Construction Process}

\begin{table*}[t]
\caption{Selected foundational benchmarks in \bench.}
\begin{adjustbox}{width=\linewidth}
\setlength{\tabcolsep}{1.5ex}
\setstretch{1.}
\label{table:base_bench}
\begin{tabular}{ccccl}
\hline
\textbf{Benchmark}                   & \textbf{Size}          & \textbf{\#Case} & \textbf{Qubits}    & \multicolumn{1}{c}{\textbf{Description}}                                                                                \\ \hline
\multirow{3}{*}{QASMBench}  & Small         & 39   & 2$\sim$10   & \multirow{3}{*}{\begin{tabular}[c]{@{}l@{}}Representative quantum algorithms of different qubit sizes\end{tabular}}                                            \\ \cline{2-4}
                            & Medium        & 24   & 11 $\sim$ 27  &                                                                                                                \\ \cline{2-4}
                            & Large         & 40   & 28 $\sim$ 127 &                                                                                                                \\ \hline
\multirow{2}{*}{VeriQBench} & Variational   & 47   & 4 $\sim$ 121  & Variational circuits, including QAOA, EfficientSU2, VQE, QCNN \\ \cline{2-5} 
                            & Combinational & 118   & 1 $\sim$ 101  & Quantum algorithms of different qubit sizes and random circuits                                                                   \\ \hline
\multicolumn{2}{c}{Total}                   & \multicolumn{3}{c}{268}                                                                                                           \\ \hline
\end{tabular}
\end{adjustbox}
\vspace{-.ex}
\end{table*}

\subsubsection{Foundation Benchmarks}\label{sec:foundation_bench}

\Cref{table:base_bench} summarizes the open-source benchmarks included in \bench, including QASMBench~\cite{QASMBench} and VeriQBench~\cite{chen2022veriqbenchbenchmarkmultipletypes}. These benchmarks cover representative quantum algorithms, variational circuits, and random circuits across diverse application scenarios. All circuits are represented in OpenQASM 2.0~\cite{openqasm2}, providing a unified format for benchmark construction and compiler option evaluation.

\begin{table*}[t]
\centering
\caption{Search space designed in \bench. The full space contains 960 configurations for each circuit.}
\label{table:search_space}
\small
\setstretch{1.}
\begin{tabularx}{\linewidth}{lX}
\toprule
\textbf{Space} & \textbf{Candidate Values} \\
\midrule
Backend &
\texttt{FakeGuadalupeV2} (16 qubits),
\texttt{FakeAuckland} (27 qubits),
\texttt{FakeBrooklynV2} (65 qubits),
\texttt{FakeBrisbane} (127 qubits) \\

Optimization level &
\texttt{0}, \texttt{1}, \texttt{2}, \texttt{3} \\

Layout &
\texttt{dense}, \texttt{sabre}, \texttt{trivial} \\

Routing &
\texttt{basic}, \texttt{lookahead}, \texttt{sabre}, \texttt{none} \\

Translation &
\texttt{synthesis}, \texttt{translator},
\texttt{ibm\_backend}, \texttt{ibm\_dynamic\_circuits},
\texttt{ibm\_fractional} \\
\bottomrule
\end{tabularx}
\vspace{-.ex}
\end{table*}

\subsubsection{Search Space}\label{sec:search_space}

\Cref{table:search_space} presents the search space and candidate values for Qiskit Compiler V2.4.1 (latest version when this work began) used for constructing \bench. We select four fake backends with different qubit sizes, choosing the device with the lowest average gate error rate in each qubit-size category. For example, \texttt{FakeBrisbane} is selected as the 127-qubit device because it has the lowest average gate error rate among the 127-qubit candidates.

For options, besides the options in \Cref{table:search_space}, Qiskit V2.4.1 also provides \texttt{approximation\_degree} ($[0,1]$), \texttt{scheduling\_method} (\texttt{None}, \texttt{alap}, or \texttt{asap}), and integer-valued \texttt{seed\_transpiler}. Since \linebreak \texttt{approximation\_degree}<1 enables approximate synthesis that may \emph{alter circuit semantics} to reduce synthesis cost, we fix it to 1.0. We set \texttt{scheduling\_method} to \texttt{None} to focus on structural compilation choices and \texttt{seed\_transpiler} to 42 for reproducibility.
For the four options in \Cref{table:search_space}, we include all available candidate values while excluding \texttt{default} choices, as they ultimately resolve to explicit methods already covered by our search space. For example, under $O_3$, the default layout and routing methods both resolve to \texttt{sabre}. Overall, each circuit has 960 candidate device-option configurations, enabling exhaustive enumeration to construct the empirical oracle in \bench.

\subsubsection{Evaluation Metric}\label{sec:search_obj}

Since directly simulating each compiled circuit to obtain its execution fidelity would incur prohibitive time and memory costs, we adopt \emph{estimated fidelity}, a widely used proxy metric in prior quantum compiler configuration selection and pass tuning studies~\cite{DAC23,mqtpredictor,zhong2026qutunerfeaturelearningguidedoptimization,tuniq}. The original estimated fidelity $F_{\mathrm{org}}$ in \Cref{eq:fidelity} is computed by multiplying the expected fidelities of all gates and measurement operations in the compiled circuit according to the hardware error information of the target device. In $F_{\mathrm{org}}$, $F(g_i)$ denotes the expected execution fidelity of gate $g_i$ on its mapped qubit(s), $F_{\mathrm{RO}}(q_j)$ denotes the expected readout fidelity of qubit $q_j$, and $|G|$ and $m$ denote the numbers of gates and measurements in the compiled circuit, respectively. When the device calibration data provides an error rate, we convert it to fidelity as $1-\mathrm{error\ rate}$.
\begin{equation}
F_{\mathrm{org}}
=
\prod_{i=1}^{|G|} F(g_i)
\prod_{j=1}^{m} F_{\mathrm{RO}}(q_j),
\qquad
F_{\mathrm{eff}}
=
\exp\left(\frac{\ln F_{\mathrm{org}}}{N_{\mathrm{input}}}\right)
=
F_{\mathrm{org}}^{1/N_{\mathrm{input}}}
\label{eq:fidelity}
\end{equation}

However, because $F_{\mathrm{org}}$ is multiplicative, circuits with more operations (gates) naturally obtain smaller values, causing cross-circuit aggregation to be biased by circuit size. Following prior work~\cite{ln_fidelity}, we normalize $F_{\mathrm{org}}$ by the number of non-barrier input operations and define the resulting \emph{effective estimated fidelity} as $F_{\mathrm{eff}}$. Since all configurations of the same circuit share the same $N_{\mathrm{input}}$, this monotonic transformation preserves their relative ordering and thus does not change the oracle configuration, while reducing the influence of circuit size.

\subsubsection{Brute-Force Search Process}\label{sec:search_process}

During the search, we enumerate all 960 device-option combinations for each circuit with a fixed seed of 42 for reproducibility. We set a 600-second timeout for each compilation to balance search cost and fidelity exploration; only five of the 268 circuits exceed this limit under the default $O_3$ configuration, indicating that it mainly filters unusually expensive cases. Finally, we select the configuration with the highest $F_{\mathrm{eff}}$ (defined in \Cref{eq:fidelity}) as the oracle, using shorter compilation time as the tie-breaker.

After 8,345.40 CPU-hours of brute-force search, we obtain oracle configurations for all 268 circuits. Parallelizing the search with 20 workers reduces the total wall-clock time to approximately 618 hours. The time cost is inflated by a small number of extremely slow circuit-configuration pairs, such as very large circuits on \texttt{FakeBrisbane} or configurations using the \texttt{lookahead} option.

The brute-force results confirm that exhaustive configuration selection is feasible for offline oracle construction but too costly for per-circuit optimization. The resulting oracle configurations provide a reference for all 268 circuits in \bench, enabling evaluation against both standard configurations and the best configuration identified over the complete search space.

\subsubsection{Level Partition}\label{sec:level_distribution}

Finally, to distinguish the optimization difficulty of cases in \bench, we partitiona the 268 cases into different levels. We first introduce a straightforward baseline, \textbf{Device-Aware O3 ($O^{DA}_3$)}, which selects the capacity-feasible device with the lowest average gate error and applies the default $O_3$ configuration (\texttt{sabre} for layout and routing and \texttt{translator} for translation). $O^{DA}_3$ reflects a natural strategy of selecting a device with better hardware quality together with a high optimization level, since the estimated fidelity~\cite{mqtpredictor,DAC23,zhong2026qutunerfeaturelearningguidedoptimization,tuniq} in \Cref{eq:fidelity} is computed from gate errors. We then quantify each circuit's optimization headroom by the oracle's relative $F_{\mathrm{eff}}$ improvement over $O^{DA}_3$ and divide the 268 circuits into three levels: 104 circuits with less than 1\% improvement as Level 1, 88 with 1--10\% as Level 2, and 76 with more than 10\% as Level 3, representing increasing optimization headroom.

We further characterize circuit complexity across the three levels. Level 2 contains relatively lighter circuits, averaging 21.8 qubits and circuit depth of 106.7, compared with 36.7 and 31,801 for Level 1, 24.5 and 39,982 for Level 3. The large average circuit depth in Levels 1 and 3 are mainly caused by a few circuits with extremely large circuit depth.

\begin{figure*}[t]
    \centering

    \begin{minipage}[t]{0.51\textwidth}
        \centering
        \includegraphics[width=\linewidth]{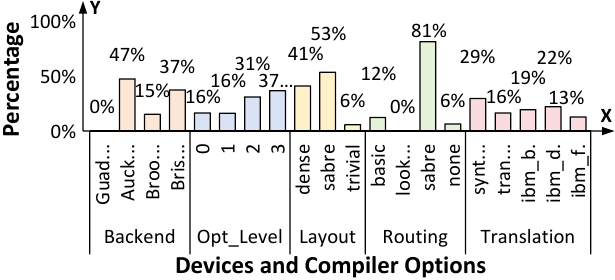}
        \captionof{figure}{The optimal configuration distribution.}
        \label{fig:distribution}
    \end{minipage}
    \hfill
    \begin{minipage}[t]{0.48\textwidth}
        \centering
        \includegraphics[width=\linewidth]{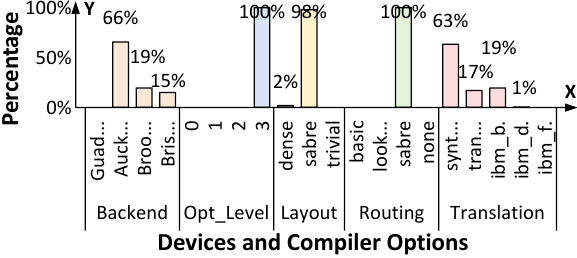}
        \captionof{figure}{GPT-5.5-selected configurations.}
        \Description{Distributions of the configuration selected by GPT-5.5.}
        \label{fig:llm_distribution}
    \end{minipage}
    \vspace{-.ex}
\end{figure*}

\subsection{Optimal Configuration Distribution}\label{sec:search_results}
We further analyze the oracle configuration distribution in \bench in \Cref{fig:distribution}. \texttt{FakeAuckland} and \texttt{FakeBrisbane} account for 47\% and 37\% of device choices, while $O_2$ and $O_3$ account for 31\% and 37\% of optimization levels. Notably, 16\% of oracle configurations use $O_0$, mainly for single-qubit-dominated circuits where stronger optimization brings little fidelity benefit and the shorter compilation time makes $O_0$ preferable. The \texttt{sabre} option is used in more than half of the cases for both layout and routing, while translation choices are more diverse. Overall, no single configuration dominates, motivating circuit-specific configuration selection.

\subsection{The Unique Value of \bench}
\bench is the first quantum compiler configuration selection benchmark that provides empirical oracle configurations for each circuit. These oracles make \bench a reusable reference for future research, enabling methods to measure both improvements over standard configurations (e.g., $O_3^{DA}$) and the remaining gap to the oracle, while supporting analyses of circuit-specific device and option choices. Although constructing the oracles is computationally expensive, this cost is incurred only once, enabling reproducible evaluation without repeating exhaustive search.

\section{An Initial Attempt at Direct Selection with LLMs}\label{sec:empirical}

Having established \bench, we next investigate LLM-based configuration selection. Unlike bottom-up methods that iteratively search the configuration space, LLMs offer a natural alternative by directly predicting configurations, as explored in classical compiler tuning~\cite{qiu2026passbypassoptimizationintentdrivenir, pan2026eccoevidencedrivencausalreasoning}. We therefore evaluate ten LLMs on \bench to characterize their effectiveness.

\subsection{Evaluation Settings}\label{sec:eva_settings}

\begin{table*}[t]
\centering
\caption{LLM-based direct configuration selection performance across three levels in \bench.}
\label{tab:llm_initial}
\setlength{\tabcolsep}{1.5ex}
\setstretch{.9}
\begin{adjustbox}{max width=\linewidth}
\begin{tabular}{llccccc|rrr}
\toprule
\multirow{2}{*}{Source}
& \multirow{2}{*}{LLM}
& \multicolumn{5}{c|}{\textbf{Effective Estimated Fidelity ($F_{eff}$ in \Cref{eq:fidelity})}}
& \multicolumn{3}{c}{\textbf{Compilation Time}} \\
\cmidrule(lr){3-7}
\cmidrule(lr){8-10}
&
& B/E/W $O^{DA}_3$
& LLM vs. $O^{DA}_3$
& $p_{\mathrm{LLM}}$
& Oracle vs. $O^{DA}_3$
& $p_{\mathrm{Oracle}}$
& $O^{DA}_3$ (s)
& LLM (s)
& Oracle (s) \\
\midrule

\multicolumn{10}{c}{\textbf{Level 1: 104 Cases}} \\
\cline{1-10}
\multirow{5}{*}{Closed-source}
& Seed-2.0-Lite    & 2/17/85 & $-43.89\%$ & $<0.001$
& \multirow{5}{*}{$0.03\%$}
& \multirow{5}{*}{$0.003$}
& \multirow{5}{*}{9.89}
& 23.28
& \multirow{5}{*}{1.50} \\
& Gemini-3.1-Flash   & 1/34/69 & $-5.23\%$  & $<0.001$ & & & & 19.62 & \\
& Gemini-3.5-Flash   & 1/34/69 & $-4.30\%$  & $<0.001$ & & & & 13.72 & \\
& GPT-5-Mini         & 1/35/68 & $-4.80\%$  & $<0.001$ & & & & 15.80 & \\ 
& GPT-5.5            & 1/83/20 & $-3.38\%$  & $<0.001$ & & & & 13.97 &  \\
\cline{1-10}
\multirow{5}{*}{Open-source}
& DeepSeek-3.2       & 0/54/50 & $-3.37\%$  & $<0.001$
& \multirow{5}{*}{$0.03\%$}
& \multirow{5}{*}{$0.003$}
& \multirow{5}{*}{9.89}
& 15.80
& \multirow{5}{*}{1.50} \\
& DeepSeek-V4-Flash  & 1/36/67 & $-7.06\%$  & $<0.001$ & & & & 16.68 & \\
& GLM-5              & 1/34/69 & $-8.38\%$  & $<0.001$ & & & & 20.83 & \\
& Qwen3.5-27B        & 1/36/67 & $-4.27\%$  & $<0.001$ & & & & 16.68 & \\
& Qwen3.5-397B-A17B  & 1/36/67 & $-12.88\%$ & $<0.001$ & & & & 16.68 & \\

\midrule
\multicolumn{10}{c}{\textbf{Level 2: 88 Cases}} \\
\cline{1-10}
\multirow{5}{*}{Closed-source}
& Seed-2.0-Lite    & 7/1/80  & $-42.90\%$ & $<0.001$
& \multirow{5}{*}{$5.83\%$}
& \multirow{5}{*}{$<0.001$}
& \multirow{5}{*}{0.09}
& 0.12
& \multirow{5}{*}{0.13} \\
& Gemini-3.1-Flash   & 6/7/75  & $-3.96\%$  & $<0.001$ & & & & 0.11 & \\
& Gemini-3.5-Flash   & 5/8/75  & $-2.87\%$  & $<0.001$ & & & & 0.10 & \\
& GPT-5-Mini         & 7/8/73  & $-2.83\%$  & $<0.001$ & & & & 0.11 & \\
& GPT-5.5            & 2/72/14 & $-3.16\%$  & $<0.001$ & & & & 0.10 & \\
\cline{1-10}
\multirow{5}{*}{Open-source}
& DeepSeek-3.2       & 3/32/53 & $-3.26\%$  & $<0.001$
& \multirow{5}{*}{$5.83\%$}
& \multirow{5}{*}{$<0.001$}
& \multirow{5}{*}{0.09}
& 6.92
& \multirow{5}{*}{0.13} \\
& DeepSeek-V4-Flash  & 7/6/75  & $-4.08\%$ & $<0.001$ & & & & 0.11 & \\
& GLM-5              & 8/5/75  & $-4.06\%$ & $<0.001$ & & & & 0.11 & \\
& Qwen3.5-27B        & 6/7/75  & $-3.24\%$ & $<0.001$ & & & & 0.10 & \\
& Qwen3.5-397B-A17B  & 9/6/73  & $-3.15\%$ & $<0.001$ & & & & 0.11 & \\

\midrule
\multicolumn{10}{c}{\textbf{Level 3: 76 Cases}} \\
\cline{1-10}
\multirow{5}{*}{Closed-source}
& Seed-2.0-Lite    & 14/18/44 & $-19.12\%$ & $<0.001$
& \multirow{5}{*}{$31.68\%$}
& \multirow{5}{*}{$<0.001$}
& \multirow{5}{*}{8.61}
& 8.83
& \multirow{5}{*}{1.06} \\
& Gemini-3.1-Flash   & 2/40/34  & $-0.59\%$ & $<0.001$ & & & & 8.61 & \\
& Gemini-3.5-Flash   & 2/40/34  & $-0.59\%$ & $<0.001$ & & & & 8.60 & \\
& GPT-5-Mini         & 2/41/33  & $-0.59\%$ & $<0.001$ & & & & 8.60 & \\
& GPT-5.5            & 4/67/5   & $0.04\%$  & $0.032$ & & & & 8.61 & \\
\cline{1-10}
\multirow{5}{*}{Open-source}
& DeepSeek-3.2       & 1/53/22  & $-0.31\%$ & $<0.001$
& \multirow{5}{*}{$31.68\%$}
& \multirow{5}{*}{$<0.001$}
& \multirow{5}{*}{8.61}
& 8.61
& \multirow{5}{*}{1.06} \\
& DeepSeek-V4-Flash  & 3/36/37  & $-0.90\%$ & $<0.001$ & & & & 8.60 & \\
& GLM-5              & 3/36/37  & $-3.92\%$ & $<0.001$ & & & & 8.60 & \\
& Qwen3.5-27B        & 2/40/34  & $-0.59\%$ & $<0.001$ & & & & 8.60 & \\
& Qwen3.5-397B-A17B  & 3/40/33  & $-0.58\%$ & $<0.001$ & & & & 8.60 & \\
\bottomrule
\end{tabular}
\end{adjustbox}
\vspace{-.ex}
\end{table*}

\noindent
\textbf{Benchmark and Evaluation Metrics.}
We evaluate LLMs on \bench using the $F_{\mathrm{eff}}$ improvement over $O^{DA}_ 3$ (defined in \Cref{sec:level_distribution}) as the primary metric, with compilation time as an efficiency metric. We measure improvement using the headroom-normalized gain, $\mathrm{Imp}(M)=\frac{F{_\mathrm{eff}}^{M}-F_{\mathrm{eff}}^{\mathrm{DA\text{-}O3}}}{1-F_{\mathrm{eff}}^{\mathrm{DA\text{-}O3}}}\times100\%$, where the denominator represents the remaining fidelity headroom of $O^{DA}_3$. Positive and negative values indicate improvement and degradation, respectively. We compute the gain per circuit and report the average within each level; oracle improvement is computed analogously.
We also report \textbf{Better/Equal/Worse} (denoted as "B/E/W") $O_3^{DA}$ as the numbers of cases where the selected configuration achieves higher, equal, or lower $F_{\mathrm{eff}}$ than $O_3^{DA}$. In addition, we report $p$-values from paired Wilcoxon signed-rank tests~\cite{significance_test} for both the selected configurations and the oracle against $O_3^{DA}$. The tests are performed on per-circuit paired \(F_{\mathrm{eff}}\) values; pairs with zero differences are excluded from the signed-rank calculation following the Wilcoxon convention. We consider cases with \(p<0.05\) statistically significant.

\noindent
\textbf{LLMs.}
We evaluate ten LLMs, including five open-source and five closed-source: GPT-5-mini~\cite{Web:gpt5}, GPT-5.5~\cite{Web:gpt55}, DeepSeek-V3.2-FP4~\cite{Web:dsv32}, DeepSeek-V4-Flash~\cite{Web:dsv4}, Gemini-3.1-Flash-Lite~\cite{Web:gemini31}, Gemini-3.5-Flash~\cite{Web:gemini}, Qwen3.5-27B~\cite{Web:qwen35}, Qwen3.5-397B-A17B~\cite{Web:qwen35}, GLM-5~\cite{Web:glm5}, and Seed-2.0-Lite~\cite{Web:doubao}.

\noindent
\textbf{Prompt Setting.}
The prompt provides the circuit features, device features, and candidate option space, asks the LLM to maximize $F_{eff}$ (\Cref{eq:fidelity}) with compilation time as a tie-breaker, and strictly requires the output configuration in the JSON format.

\noindent
\textbf{Implementation Details.}
We perform inference for all LLMs using the Python APIs provided by OpenRouter~\cite{Web:openrouter}. For all models, we set the maximum output length to 1,024 tokens, the temperature to 0.0, the seed to 42, and use the default reasoning setting. All experiments are conducted on a server equipped with a 64-core Intel Xeon Gold CPU running Ubuntu 24.04.3 LTS.

\subsection{Evaluation Results}\label{sec:empirical_results}

As shown in \Cref{tab:llm_initial}, we report the average $F_{\mathrm{eff}}$ improvement over $O_3^{DA}$ and compilation time of ten LLMs on \bench, together with the oracle. Vanilla LLMs generally underperform $O_3^{DA}$ across all three levels and remain far from the oracle, showing that general-purpose LLM reasoning alone is ineffective for Qiskit configuration selection. Moreover, Level 2 shows noticeably shorter compilation time than Levels 1 and 3; as discussed in \Cref{sec:level_distribution}, this mainly reflects its generally shallower circuits, while a few extremely large circuits inflate the mean times of Levels 1 and 3. Paired significance tests further show that all LLM--$O_3^{DA}$ differences are statistically significant ($p<0.05$). We further analyze the causes of this performance gap in the next subsection.

\begin{table*}[t]
\centering
\caption{Aggressiveness scores assigned to device features and compiler-option values.}
\label{table:aggressive_score}
\setstretch{1}
\setlength{\tabcolsep}{1ex}

\begin{adjustbox}{width=\linewidth}
\begin{tabular}{cccc|cccc}
\hline
\multicolumn{4}{c|}{\textbf{Device Dimensions}} &
\multicolumn{4}{c}{\textbf{Compiler-Option Dimensions}} \\
\hline
\textbf{Device} &
\textbf{Qubit Size} &
\textbf{Topology} &
\textbf{Gate Error} &
\textbf{Optimization Level} &
\textbf{Layout} &
\textbf{Routing} &
\textbf{Translation} \\
\hline

Score &
$\operatorname{rank}_{\uparrow}(Q_d)$ &
$\operatorname{rank}_{\uparrow}(T_d)$ &
$\operatorname{rank}_{\downarrow}(E_d)$ &
O0: 0 & trivial: 0 & none: 0 & translator: 0 \\

FakeGuadalupeV2 &
0 & 1 & 1 &
O1: 1 & dense: 1 & basic: 1 & ibm\_backend: 1 \\

FakeAuckland &
1 & 2 & 3 &
O2: 2 & sabre: 2 & lookahead: 2 & ibm\_dynamic: 1 \\

FakeBrooklynV2 &
2 & 3 & 2 &
O3: 3 & & sabre: 3 & ibm\_fractional: 1 \\

FakeBrisbane &
3 & 0 & 0 &
& & & synthesis: 2 \\

\hline
\end{tabular}
\end{adjustbox}
\vspace{-.ex}
\end{table*}

\subsection{Analysis}\label{sec:empirical_analysis}

\subsubsection{LLM-Selected Configuration Distribution}\label{sec:option_analysis}

We further analyze GPT-5.5, the best-performing LLM on average in \Cref{tab:llm_initial}, and compare its predicted device-option distribution with the oracle in \Cref{fig:distribution}. We quantitatively analyze all LLMs in the next subsection. As shown in \Cref{fig:llm_distribution}, GPT-5.5 selects \texttt{FakeAuckland} about 20\% more often and \texttt{FakeBrisbane} about 20\% less often. It also favors more aggressive compiler options, including $O_3$, \texttt{sabre} for layout and routing, and \texttt{synthesis} for translation. We refer to this tendency as \textbf{\emph{configuration aggressiveness}}, i.e., a preference for more complex and transformation-intensive compilation strategies. However, the oracle distribution shows that less aggressive choices, such as $O_2$, \texttt{dense} layout, and \texttt{basic} routing, can also achieve higher fidelity and shorter compilation time. This suggests that vanilla LLMs over-rely on conventionally "strong" settings rather than adapting to circuit- and device-specific characteristics.

\subsubsection{Configuration Aggressiveness Metric}

\begin{figure}
    \centering
    \includegraphics[width=\linewidth]{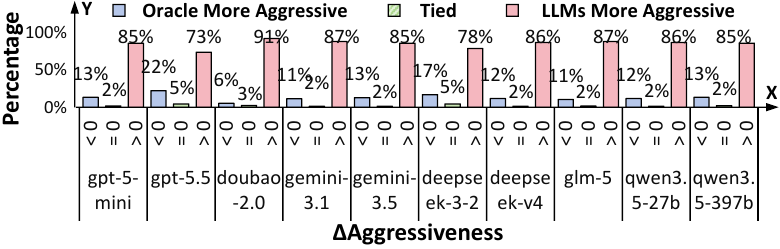}
    \caption{The $\Delta_{\mathrm{Aggressiveness}}$ metric distribution across all ten LLMs.}
    \Description{xx}
    \label{fig:metric_distribution}
    \vspace{-.ex}
\end{figure}

To better characterize the aggressiveness of different configurations, we introduce an \emph{aggressiveness metric}. As shown in \Cref{table:aggressive_score}, we assign an aggressiveness score to each candidate device and compiler-option value. For devices, aggressiveness captures the tendency to select a stronger or larger hardware target rather than compilation intensity itself. We consider three hardware characteristics: qubit capacity, topology connectivity, and gate error. A device with more qubits, higher connectivity, and lower gate error represents a stronger hardware choice and is therefore assigned a higher aggressiveness score. For compiler options, we assign scores according to  the complexity and transformation intensity of each option, as detailed in \Cref{sec:option_analysis}.

\begin{equation}
\small
\begin{aligned}
\mathrm{Agg}(o)=\;&
\operatorname{round}\left(
\frac{
s_{\mathrm{q}}(o_{\mathrm{dev}})
+s_{\mathrm{t}}(o_{\mathrm{dev}})
+s_{\mathrm{g}}(o_{\mathrm{dev}})
}{3}
\right) \\
&+s_{\mathrm{opt}}(o_{\mathrm{opt}})
+s_{\mathrm{lay}}(o_{\mathrm{lay}}) \\
&+s_{\mathrm{rout}}(o_{\mathrm{rout}})
+s_{\mathrm{trans}}(o_{\mathrm{trans}})
\end{aligned}
\label{eq:aggressive_score}
\end{equation}

The overall aggressiveness score of a configuration $o$ is calculated using \Cref{eq:aggressive_score}, where the three device-related scores are averaged so that device contributes as one component. A larger $\mathrm{Agg}(o)$ indicates a more aggressive selected quantum compiler configuration.

We further compare each LLM-selected configuration with its oracle counterpart using  the $\Delta_{\mathrm{Aggressiveness}}$, defined as \emph{the LLM aggressiveness score \textbf{minus} the oracle score}. A positive value indicates that the LLM selects a more aggressive configuration, while a negative value indicates that the oracle is more aggressive. As shown in \Cref{fig:metric_distribution}, the \textcolor{pink}{\textit{\textbf{pink}}} regions corresponding to $\Delta_{\mathrm{Aggressiveness}}>0$ dominate across all ten LLMs, confirming a clear \textbf{over-aggressiveness bias}: vanilla LLMs tend to favor more aggressive quantum compiler configurations than the oracle.
\section{Oracle-Guided Rule Distillation for Top-Down Configuration Selection}\label{sec:rule}

\begin{figure*}[t]
    \centering
    \includegraphics[width=0.95\linewidth]{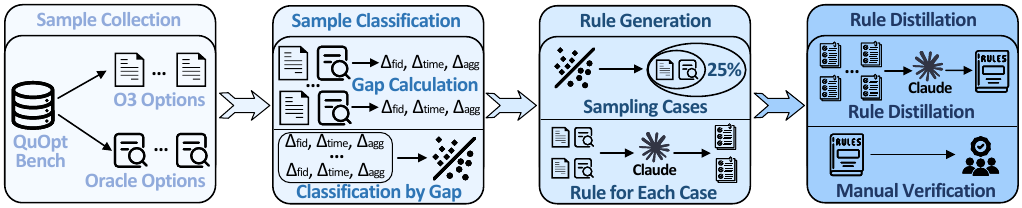}
    \caption{The quantum compiler configuration selection rule distillation workflow.}
    \Description{xx}
    \label{fig:rule}
    \vspace{-.ex}
\end{figure*}

In \Cref{sec:empirical}, vanilla LLMs fail to achieve effective direct configuration selection. This suggests that effective direct selection requires explicit knowledge of how configuration choices relate to circuit and device characteristics. We therefore design a four-step rule-distillation workflow to extract transferable configuration-selection rules from oracle decisions, as shown in \Cref{fig:rule}.

\subsection{Sample Collection and Classification}\label{sec:sample_classification}

Initially, we collect all 268 cases from \bench, using both $O_3^{DA}$ and the oracle configurations as comparison references. To distill comprehensive and generalizable rules, the sampled cases should cover diverse differences between these two configurations. Therefore, before sampling, we classify all cases according to their effective estimated fidelity, compilation-time, and aggressiveness score gaps, ensuring that the selected samples represent diverse optimization scenarios.

\begin{table*}[t]
\centering
\caption{Gap-based sample classification rules for rule distillation. We set the fidelity-gap threshold $\tau_{F_{eff}}=0.01$, the time-overhead threshold $\tau_T=1.5$, and the aggressiveness score threshold $\tau_{Agg}=2$.} 
\label{table:gap_classification} 
\begin{adjustbox}{max width=\linewidth} 
\begin{tabular}{lll} \hline 
Category & Rule & Description \\ \hline 
Fidelity-loss / $O_3^{DA}$-more-aggressive & $\Delta_{F_{eff}}>\tau_{F_{eff}} \land \Delta_{Agg}>\tau_{Agg}$ & Fidelity loss when $O_3^{DA}$ is more aggressive than the oracle \\ 
Fidelity-loss / Oracle-more-aggressive & $\Delta_{F_{eff}}>\tau_{F_{eff}} \land \Delta_{Agg}<-\tau_{Agg}$ & Fidelity loss when the oracle is more aggressive than $O_3^{DA}$ \\ 
Fidelity-loss / Similar-aggressiveness & $\Delta_{F_{eff}}>\tau_{F_{eff}} \land -\tau_{Agg}\le\Delta_{Agg}\le\tau_{Agg}$ & Fidelity loss with similar aggressiveness scores \\ 
Time-overhead / $O_3^{DA}$-more-aggressive & $\Delta_{F_{eff}}\le\tau_{F_{eff}} \land \Delta_T>\tau_T \land \Delta_{Agg}>\tau_{Agg}$ & Time overhead when $O_3^{DA}$ is more aggressive than the oracle \\ 
Time-overhead / Oracle-more-aggressive & $\Delta_{F_{eff}}\le\tau_{F_{eff}} \land \Delta_T>\tau_T \land \Delta_{Agg}<-\tau_{Agg}$ & Time overhead when the oracle is more aggressive than $O_3^{DA}$ \\ 
Time-overhead / Similar-aggressiveness & $\Delta_{F_{eff}}\le\tau_{F_{eff}} \land \Delta_T>\tau_T \land -\tau_{Agg}\le\Delta_{Agg}\le\tau_{Agg}$ & Time overhead with similar aggressiveness scores \\ 
Near-optimal & $\Delta_{F_{eff}}\le\tau_{F_{eff}} \land \Delta_T\le\tau_T$ & Close to the oracle in both fidelity and compilation time \\ \hline 
\end{tabular} 
\end{adjustbox} 
\vspace{-.ex}
\end{table*}

We first define the gaps in effective estimated fidelity ($\Delta_{F_{eff}}$), compilation time ($\Delta_{T}$), and configuration aggressiveness ($\Delta_{Agg}$) in \Cref{eq:clf_gap}.

\begin{equation}
\label{eq:clf_gap}
\begin{aligned}
\Delta_{F_{eff}}
&= {F_{eff}}^{\mathrm{Oracle}}
 - {F_{eff}}^{\mathrm{O_3^{DA}}}, \\
\Delta_T
&= {T_{\mathrm{O_3^{DA}}}} \,/\, {T_{\mathrm{Oracle}}}, \\
\Delta_{Agg}
&= \mathrm{Agg}_{\mathrm{O_3^{DA}}}
 - \mathrm{Agg}_{\mathrm{Oracle}}.
\end{aligned}
\end{equation}

These metrics characterize the differences between the oracle and $O_3^{DA}$ from three complementary perspectives: compilation quality, compilation cost, and configuration aggressiveness. A higher $F_{\mathrm{eff}}$ indicates better effective estimated fidelity, a lower compilation time indicates lower compilation overhead, and the aggressiveness score reflects the extent to which a configuration adopts more aggressive compilation strategies.
Here, $F_{eff}^{\mathrm{Oracle}}$ and $F_{eff}^{O_3^{DA}}$ denote the effective estimated fidelities of the oracle and $O_3^{DA}$ configurations, respectively; $T_{\mathrm{Oracle}}$ and $T_{O_3^{DA}}$ denote their compilation time overhead; and $\mathrm{Agg}_{\mathrm{Oracle}}$ and $\mathrm{Agg}_{O_3^{DA}}$ denote their aggressiveness scores computed using \Cref{eq:aggressive_score}. Accordingly, $\Delta_{F_{eff}} \ge 0$ always holds; $\Delta_T>1$ indicates that $O_3^{DA}$ requires longer compilation time; and $\Delta_{\mathrm{Agg}}>0$ indicates that $O_3^{DA}$ is more aggressive than the oracle.

Next, we define the classification rules in \Cref{table:gap_classification}. Based on \Cref{eq:clf_gap}, we set the thresholds to $\tau_{F_{eff}}=0.01$ for effective estimated fidelity, $\tau_T=1.5$ for compilation-time ratio, and $\tau_{Agg}=2$ for aggressiveness gap. These values are chosen empirically: a fidelity gap above $0.01$ indicates noticeable quality difference, a time ratio above $1.5$ indicates clear compilation overhead, and an aggressiveness gap above $2$ reflects a substantial difference in configuration aggressiveness.

We then divide the cases into seven categories based on fidelity loss, compilation-time overhead, and aggressiveness differences. For both fidelity-loss cases and cases with similar fidelity but substantial time overhead, we distinguish whether $O_3^{DA}$ is more aggressive, less aggressive, or similar to the oracle; the remaining cases are classified as \emph{near-optimal}. This classification rule follows our fidelity-first, compilation-time-second evaluation priority, as used in the construction process of  \bench (\Cref{sec:search_process}).

Among the seven categories in \Cref{table:gap_classification}, only four contain cases in \bench. We then randomly sample \textbf{25\%} of the cases from each non-empty category, yielding \textbf{68 cases}, while reserving the remaining 200 exclusively for evaluation to reduce overfitting.
Among the sampled cases, 52.94\% are \texttt{Fidelity\_loss/$O_3^{DA}$} more aggressive, 27.94\% are \texttt{Near-optimal}, 11.76\% are \texttt{Time-overhead/$O_3^{DA}$ more aggressive}, and 7.35\% are \texttt{Fidelity\_loss/Oracle} more aggressive. Overall, 64.71\% of the sampled cases have $O_3^{DA}$ more aggressive than the oracle, suggesting that over-aggressive compilation is a major source of suboptimality. Meanwhile, the 27.94\% \texttt{Near-optimal} cases show that $O_3^{DA}$ remains effective for a substantial subset of circuits. The other three categories are empty in \bench: no cases exhibit similar aggressiveness with substantial fidelity loss or time overhead, and no \texttt{Time-overhead/Oracle-more-aggressive} cases are observed.

\subsection{Rule Generation and Distillation}\label{sec:rule_gen}

For each sampled case, we provide Claude-Opus-5~\cite{Web:claude} with the $O_3^{DA}$ and oracle configurations, circuit features, and device features. We ask Claude to analyze why the oracle is preferable, which fidelity or compilation-time differences are caused by $O_3^{DA}$, and which circuit feature, device, or option are most relevant. Based on this analysis,  Claude-Opus-5 extracts one case-level rule with explicit applicability conditions, specifying under what circuit and device features a particular device or compiler option should be preferred. Through this process, we obtain 68 case-level rules.

We use  Claude-Opus-5 because it was the latest Claude model with strong reasoning capability when we conducted rule extraction. We further evaluate the impact of the distilling LLM through an ablation study with two alternative LLMs in \Cref{sec:aba_study}.

Finally, we provide all case-level rules to  Claude-Opus-5 and ask it to consolidate recurring patterns into a compact rule set, while also identifying the supporting cases for each distilled rule. We then manually inspect the distilled rules to verify that each rule is consistent with the circuit and device features of its supporting cases, remove redundant or overly case-specific statements, and check the overall consistency of the rule set. The final rule set is shown in \Cref{tab:distilled_rules}.

\begin{table*}[t]
\centering
\footnotesize
\caption{Distilled rules for quantum compiler configuration selection.}
\label{tab:distilled_rules}
\setstretch{1}
\renewcommand{\arraystretch}{1.08}
\setlength{\tabcolsep}{4pt}

\begin{tabularx}{\textwidth}{|p{0.04\textwidth} |X|}
\hline
\textbf{Index} & \textbf{Specific Rules} \\
\hline

\textbf{R1} & For 2--11-qubit circuits with backend occupancy $\mathrm{Occ}(c,b)=N_q(c)/N_q(b)\leq0.4$, favor wider feasible backends with higher mean $T_1/T_2$ and \texttt{layout=dense}. At higher occupancy, prefer backends closer to the circuit width. \\ \hline

\textbf{R2} & For 11--39-qubit circuits with occupancy $\leq0.35$, low communication (\texttt{program\_communication}$\leq0.15$), and \texttt{parallelism}$\leq0.15$, favor wider backends with higher mean $T_1/T_2$ and use O2--O3 with SABRE layout/routing. We regard \texttt{program\_communication}$\geq0.5$ as high communication; in such cases, prioritize backend connectivity, measured by \texttt{avg\_degree} and \texttt{edge\_density}. \\ \hline

\textbf{R3} & For 2--8-qubit circuits with \texttt{entanglement\_ratio}$\leq0.25$ and \texttt{parallelism}$\leq0.15$, prefer O0--O1, \texttt{trivial}/\texttt{dense} layout, \texttt{basic} routing, and non-synthesis translation. Here, non-synthesis translation refers to \texttt{translator}, \texttt{ibm\_backend}, \texttt{ibm\_dynamic\_circuits}, or \texttt{ibm\_fractional}, as applicable. \\ \hline

\textbf{R4} & For shallow 1--8-qubit circuits with \texttt{num\_2q\_gates}=0 and \texttt{num\_multi\_qubit\_gates}=0, use O0--O1 with \texttt{layout=dense}; routing is unnecessary, and aggressive SABRE-based mapping provides little benefit. \\ \hline

\textbf{R5} & For near-native circuits, i.e., circuits without $>2$-qubit composite blocks whose source operations are already supported or directly lowered to the backend basis, prefer O2--O3 with SABRE layout/routing and non-synthesis translation. Use \texttt{synthesis} when unsupported or composite multi-qubit operations require substantial decomposition. \\ \hline

\textbf{R6} & For circuits dominated by higher-arity operations, where $N_{>2q}=\texttt{num\_multi\_qubit\_gates}-\texttt{num\_2q\_gates}$ exceeds \texttt{num\_2q\_gates}, prefer O1--O2 with SABRE routing and a structure-matched non-synthesis translation method. Use O3/synthesis only when substantial optimization opportunity is predicted. \\ \hline
\textbf{R7} & For circuits with $\geq55$ qubits and occupancy $>0.4$, prefer O0--O2 with \texttt{trivial}/\texttt{dense} layout and SABRE routing. Retain O3/synthesis when the circuit contains unsupported parameterized or controlled operations that require substantial synthesis. \\ \hline

\textbf{R8} & For extremely deep and high-communication circuits (\texttt{depth}$\geq10^5$, \texttt{parallelism}$\leq0.05$), prefer O0--O2, dense layout, SABRE routing, and non-synthesis translation. Avoid heavy optimization when the circuit is already near the backend basis. \\ \hline

\textbf{R9} & For $\geq30$-qubit, low-communication circuits on sparse backends, prefer O0--O2, SABRE layout/routing, and non-synthesis translation when near-native; use O3/synthesis for substantial non-native or redundant structure. \\ \hline

\textbf{R10} & For circuits containing \texttt{reset} or conditional-control operations, prefer \texttt{ibm\_dynamic\_circuits}; otherwise prefer \texttt{translator} or \texttt{ibm\_backend} over generic synthesis when the gate set is already near-native. \\  \hline

\textbf{R11} & For near-full-device, serial entangling circuits with occupancy $>0.9$, \texttt{entanglement\_ratio}$\geq0.8$, \texttt{critical\_depth}$\geq0.8$, and \texttt{parallelism}$\leq0.05$, prefer \texttt{basic}/SABRE routing with non-synthesis translation.\\ \hline

\textbf{R12} & For high-communication circuits, prioritize feasible backends with higher connectivity and lower two-qubit error, together with SABRE routing. At low occupancy, use \texttt{dense} layout; use synthesis only for non-native operations. \\ \hline

\textbf{R13} & For 5--16-qubit circuits with occupancy $\leq0.4$, \texttt{program\_communication}$\leq0.4$, and \texttt{parallelism}$\leq0.15$, prefer O0--O2 while retaining SABRE layout/routing. Escalate to O3 only when the predicted optimization opportunity is sufficiently large.\\ \hline
\end{tabularx}
\vspace{-.ex}
\end{table*}

Distilled rules in \Cref{tab:distilled_rules} reveal three common patterns. First, device selection depends on circuit width and communication: low-occupancy circuits can benefit from wider devices, while high-communication circuits favor better connectivity and lower two-qubit error. Second, aggressive compilation is not universally beneficial; simple, serial, or near-full-device circuits often prefer lower optimization levels, whereas $O_3$ is mainly useful when sufficient optimization opportunity exists. Third, translation should match circuit structure: near-native circuits generally favor lightweight non-synthesis translation, while synthesis is more useful for unsupported or higher-arity operations. Overall, these patterns show that effective configuration selection requires circuit- and device-specific adaptation rather than a fixed aggressive strategy.
\section{\system: A Rule-Guided ML Framework for Top-Down Configuration Selection}\label{sec:qutop}

The distilled rules in \Cref{sec:rule} provide transferable knowledge for top-down quantum compiler configuration selection. A straightforward way to apply these rules is to include them in LLM prompts. However, LLM-based selection introduces substantial inference overhead, as LLM reasoning can take up to 82.46 seconds for a single circuit in our evaluation in \Cref{tab:llm_initial}.

\begin{figure}
    \centering
    \includegraphics[width=\linewidth]{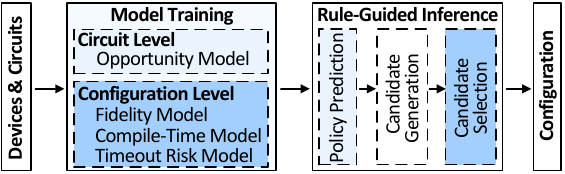}
    \caption{The workflow of \system.}
    \label{fig:heurstic}
    \vspace{-.ex}
\end{figure}

Such latency limits the practicality of using LLMs directly as online configuration selectors. To provide a fast realization of the distilled top-down rules, we develop \system, a lightweight rule-guided ML framework for configuration selection. Instead of asking an LLM to reason over all candidate configurations at inference time, \system uses lightweight predictive models to estimate the optimization opportunity and score the candidate configurations, while the distilled rules constrain candidate generation. The predicted scores are then used to efficiently select the most promising configuration as the final output.

As shown in \Cref{fig:heurstic}, \system consists of two phases: \emph{predictive model training} and \emph{rule-guided inference}. During training, we train four lightweight ML models to predict complementary signals, including circuit-level optimization opportunity, configuration-level fidelity gain, compilation-time overhead, and timeout risk. During inference, these predictions are first used to estimate each circuit's optimization headroom over $O_3^{DA}$ and assign it to an appropriate selection policy. \system then combines the selected policy with the distilled rules to generate candidate configurations and scores all candidates using a unified objective. The highest scored configuration is finally selected. We provide the full algorithm of \system in \Cref{alg:rule_guided_predictive_search}.

\begin{algorithm}[ht]
\caption{A Rule-Guided ML Framework for Top-Down Configuration Selection} 
\label{alg:rule_guided_predictive_search}
\setstretch{.7}
\scriptsize

\KwIn{Training dataset $\mathcal{D}_{train}$, test set $\mathcal{D}_{test}$, candidate devices $\mathcal{B}$, option space $\mathcal{O}$, distilled rules $\mathcal{R}$, three gate thresholds $\tau_{\mathrm{safe}},\tau_{\mathrm{seek}},\tau_{\mathrm{risk}}$, and two scoring weights $\lambda_r,\lambda_t$.}
\KwOut{Selected device-option configuration $a_c^*$ for each $c\in\mathcal{D}_{test}$.}

\BlankLine
\tcp{\textbf{Phase I: Predictive Model Training}}

$\mathcal{D}_{opp}\leftarrow\varnothing$
\tcp*{Circuit-Level: Circuit features $\rightarrow$ Maximum fidelity improvement over $O_3^{DA}$}

$\mathcal{D}_{cfg}\leftarrow\varnothing$
\tcp*{Cfg-Level: (circuit features,device,option) $\rightarrow$ fidelity gain, timeout, and time cost}

\ForEach{$c\in\mathcal{D}_{train}$}{

$\mathcal{B}_c\leftarrow\mathrm{FilterFeasibleDevices}(\mathcal{B},c)$; \quad
$\mathcal{A}_c\leftarrow\mathcal{B}_c\times\mathcal{O}$
\tcp*{Construct all feasible configurations}

$a_{O_3^{DA}}\leftarrow\mathrm{GetO3Cfg}(c)$; \quad
$(f_{O_3^{DA}},t_{O_3^{DA}})\leftarrow\mathrm{Observe}(c,a_{O_3^{DA}})$
\tcp*{Use $O_3^{DA}$ as the reference configuration}

\ForEach{$a=(b,o)\in\mathcal{A}_c$}{

    $(f_a,t_a,r_a)\leftarrow\mathrm{Observe}(c,a)$
    \tcp*{$r_a=1$ if compilation times out, and $0$ otherwise}

    $\delta_{f_a}\leftarrow\mathrm{Fidelity\_Gain}(f_a,f_{O_3^{DA}})$; \quad
    $\delta_{t_a}\leftarrow\mathrm{Time\_Overhead}(t_a,t_{O_3^{DA}})$
    
    $\mathcal{D}_{cfg}\leftarrow\mathcal{D}_{cfg}
    \cup\{(\mathrm{Concat}(c,b,o),\delta_{f_a},r_a,\delta_{t_a})\}$
    \tcp*{Store candidate-level training samples}
}

$o_c\leftarrow\max_{a\in\mathcal{A}_c}\delta_{f_a}$
\tcp*{Define case-level optimization opportunity as the best achievable gain}

$\mathcal{D}_{opp}\leftarrow\mathcal{D}_{opp}
\cup\{(\mathrm{Concat}(c,{b_{O_3^{DA}}},o_{O_3^{DA}}),o_c)\}$
\tcp*{Store one opportunity-training sample for circuit $c$}

}

\tcp{Train models to predict the circuit-level maximum fidelity gain over $O_3^{DA}$ ($M_\mathrm{opp}$), and configuration-level gain ($M_\mathrm{gain}$), timeout risk ($M_\mathrm{risk}$), and compilation-time cost ($M_\mathrm{time}$)}

$M_{\mathrm{opp}}\leftarrow\mathrm{Train}(\mathcal{D}_{opp})$; \quad
$M_{\mathrm{gain}}\leftarrow\mathrm{Train}(\mathcal{D}_{cfg},\delta_f)$; \quad
$M_{\mathrm{risk}}\leftarrow\mathrm{Train}(\mathcal{D}_{cfg},r)$; \quad
$M_{\mathrm{time}}\leftarrow\mathrm{Train}(\mathcal{D}_{cfg},\delta_t)$

\BlankLine
\tcp{\textbf{Phase II: Rule-Guided Inference}}

\ForEach{$c\in\mathcal{D}_{test}$}{

    $\mathcal{B}_c\leftarrow\mathrm{FilterFeasibleDevices}(\mathcal{B},c)$; \quad
    $a_{O_3^{DA}}\leftarrow\mathrm{GetO3Cfg}(c)$; \quad
    $x_{O_3^{DA}}\leftarrow\mathrm{Concat}(c,{b_{O_3^{DA}}},o_{O_3^{DA}})$
    
    $\hat{o}_c\leftarrow M_{\mathrm{opp}}(x_{O_3^{DA}})$; \quad
    $\hat{r}_{O_3^{DA}}\leftarrow M_{\mathrm{risk}}(x_{O_3^{DA}})$
    \tcp*{Predict optimization opportunity and timeout risk of $O_3^{DA}$}

\BlankLine
\tcp{\textbf{Policy gate: KEEP preserves $O_3^{DA}$, SAFE/SEEK allows local/broader modifications}}

\eIf{$\hat{r}_{O_3^{DA}}\ge\tau_{\mathrm{risk}}$}{
    $p_c\leftarrow\textsc{SEEK}$ if $\hat{o}_c\ge\tau_{\mathrm{seek}}$, otherwise $\textsc{SAFE}$
}{
    $p_c\leftarrow\textsc{KEEP}$ if $\hat{o}_c<\tau_{\mathrm{safe}}$; \quad
    $\textsc{SAFE}$ if $\tau_{\mathrm{safe}}\le\hat{o}_c<\tau_{\mathrm{seek}}$; \quad
    $\textsc{SEEK}$ otherwise
}

    \BlankLine
    \tcp{\textbf{Rule-guided candidate generation: rules determine how to modify $O_3^{DA}$}}

    $\mathcal{C}_c\leftarrow\{a_{O_3^{DA}}\}$

    \If{$p_c\neq\textsc{KEEP}$}{

        \ForEach{$b\in\mathcal{B}_c$}{

            $\mathcal{O}^{\mathcal{R}}_{c,b}\leftarrow
            \mathrm{ApplyRules}(c,b,\mathcal{O},\mathcal{R})$
            \tcp*{Keep option values supported by matched rules}

            \eIf{$p_c=\textsc{SAFE}$}{
                $\mathcal{O}'_{c,b}\leftarrow
                \mathrm{LocalSearch}_{50\%}(\mathcal{O}^{\mathcal{R}}_{c,b},o_{O_3^{DA}})$
                \tcp*{Allow changes to at most 50\% of option dimensions}
            }{
                $\mathcal{O}'_{c,b}\leftarrow\mathcal{O}^{\mathcal{R}}_{c,b}$
                \tcp*{Allow device switch and all rule-supported option changes}
            }

            $\mathcal{C}_c\leftarrow\mathcal{C}_c
            \cup\{(b,o)\mid o\in\mathcal{O}'_{c,b}\}$
        }
    }

    \BlankLine
    \tcp{\textbf{Candidate scoring}}

    \ForEach{$a=(b,o)\in\mathcal{C}_c$}{

        $x_a\leftarrow\mathrm{Concat}(c,b,o)$; \quad
        $\hat{g}_a\leftarrow M_{\mathrm{gain}}(x_a)$; \quad
        $\hat{r}_a\leftarrow M_{\mathrm{risk}}(x_a)$; \quad
        $\hat{h}_a\leftarrow M_{\mathrm{time}}(x_a)$

        $s_a\leftarrow\hat{g}_a-\lambda_r\hat{r}_a-\lambda_t\hat{h}_a$
        \tcp*{Fidelity is primary; timeout risk and compilation time are penalties}
    }

    $a_c^*\leftarrow\arg\max_{a\in\mathcal{C}_c}s_a$
    \tcp*{Select the device-option configuration with the highest score}
}

\Return{$\{a_c^*\mid c\in\mathcal{D}_{test}\}$}

\end{algorithm}

\subsection{Predictive Model Training}\label{sec:predictive_training}

We use the 68 rule-distillation cases as the model-development set $\mathcal{D}_{\mathrm{train}}$, with an 80\%/20\% split within each category in \Cref{table:gap_classification} for model training and validation (54/14 cases). Since we record the outcomes of all enumerated configurations when constructing \bench, these cases provide complete performance labels over the entire configuration space. The remaining 200 cases are reserved as the test set $\mathcal{D}_{\mathrm{test}}$ for evaluation.

For each training circuit $c$ in $\mathcal{D}_{train}$, we filter infeasible devices and construct all feasible configurations $\mathcal{A}c=\mathcal{B}c\times\mathcal{O}$ (line 4 in \Cref{alg:rule_guided_predictive_search}), using $O_3^{DA}$ as the reference (line 5). For each configuration, we compute its fidelity gain and compilation-time overhead relative to $O_3^{DA}$, and record whether the compilation times out. We then build a circuit-level opportunity dataset $\mathcal{D}_{opp}$ (lines 6--11), where the optimization opportunity $o_c$ represents the maximum achievable fidelity gain over $O_3^{DA}$. The motivation for measuring optimization opportunity comes from the three levels identified in \bench: for some circuits, $O_3^{DA}$ is already close to the oracle, while others still have substantial fidelity headroom. Therefore, $o_c$ estimates whether searching beyond $O_3^{DA}$ is worthwhile and how aggressively the configuration space should be explored.
The second is a \emph{configuration-level dataset} $\mathcal{D}_{cfg}$, which maps circuit, device, and option to the observed fidelity gain, timeout risk, and compilation-time overhead of each candidate configuration. The timeout risk captures a configuration's likelihood of compilation timeout, particularly for large circuits.

We train four \emph{Extremely Randomized Trees} (Extra Trees)~\cite{ertree} models (line 12): $M_{\mathrm{opp}}$, $M_{\mathrm{gain}}$, $M_{\mathrm{risk}}$, and $M_{\mathrm{time}}$, predicting optimization opportunity, fidelity gain, risk, and compilation-time overhead, respectively. Circuit features, devices, and options are vectorized as model inputs. Each model uses 250 trees, a minimum leaf size of 2, and seed 42. We choose Extra Trees because it efficiently captures non-linear interactions in structured tabular features while providing much lower inference overhead than neural or LLM-based predictors.

\subsection{Rule-Guided Inference} \label{sec:rg_inf}

\subsubsection{Policy Prediction}\label{sec:policy}

As shown in \Cref{tab:llm_initial}, the optimization opportunity over $O_3^{DA}$ varies substantially across circuits, motivating an adaptive policy rather than a uniform modification scope. Near-optimal cases should preserve $O_3^{DA}$ to avoid unnecessary modifications, whereas circuits with larger optimization opportunities require broader configuration changes. Accordingly, we introduce three policy modes with different modification scopes.

For a new circuit $c$, \system predicts its optimization opportunity $\hat{o}_c$ over $O_3^{DA}$ using $M{_\mathrm{opp}}$ and its timeout risk $\hat{r}_{O3}$ using $M{_\mathrm{risk}}$ (lines 13--15). Based on these predictions, the policy gate assigns one of three modes to $c$: \textsc{KEEP} retains $O_3^{DA}$ when little improvement is predicted; \textsc{SAFE} allows constraint modifications to device and options for moderate headroom; and \textsc{SEEK} enables broader device and option changes when substantial improvement is predicted.
As shown in lines 16--19 of \Cref{alg:rule_guided_predictive_search}, the gate is controlled by $\tau_{\mathrm{safe}}$, $\tau_{\mathrm{seek}}$, and $\tau_{\mathrm{risk}}$, avoiding unnecessary modification for near-optimal cases while expanding the candidate scope for cases with larger predicted headroom.

\subsubsection{Rule-Guided Candidate Generation} \label{sec:rule_candidate}
For circuits assigned to \textsc{SAFE} or \textsc{SEEK}, \system uses the distilled rules in \Cref{tab:distilled_rules} to generate \textbf{\emph{all}} rule-supported candidate configurations (lines 20--28), which are then passed to the subsequent scoring stage for selection.
For each capacity-feasible device $b$, \system matches the circuit and device features against the applicability conditions of the rules and retains only the option values supported by the matched rules, forming a rule-guided option space $\mathcal{O}^{\mathcal{R}}_{c,b}$ (line 23). In this way, oracle-derived knowledge is transferred to new circuits by narrowing the modification toward configurations that are more likely to be suitable.

The \textsc{SAFE} and \textsc{SEEK} modes differ in candidate scope. Under SAFE (lines 24--25), QuTop performs a local modification around $O_3^{DA}$ and allows at most 50\% option changes, while still permitting device switches supported by the rules. This reduces the risk of unnecessary large modifications when $O_3^{DA}$ is already reasonably strong. Under \textsc{SEEK}, all rule-supported device and option changes are considered, enabling broader exploration for circuits with larger predicted optimization headroom.
Importantly, the distilled rules do not directly determine the final configuration. Instead, they act as \emph{top-down selection priors} that define promising regions of the configuration space, while the predictive models subsequently evaluate and rank the generated candidates. Finally, all rule-guided device-option configurations are collected as candidates for subsequent scoring (line 28).

\subsubsection{Candidate Scoring and Selection} \label{sec:candidate_scoring}

After generating the candidate set $\mathcal{C}_c$, \system uses the three configuration-level models to predict each candidate's fidelity gain, timeout risk, and compilation-time overhead, and combines them into a unified score (lines 29--32 in \Cref{alg:rule_guided_predictive_search}). Fidelity remains the primary objective, while timeout risk and compilation-time overhead are used as penalties. \system then selects the candidate with the highest score as the final configuration (line 32).

By scoring only a rule-pruned candidate set with lightweight models, \system avoids both repeated online compilation in bottom-up search and the high reasoning latency of LLM-based top-down selection, providing a fast realization of the proposed top-down configuration selection paradigm.

\section{Evaluation}

The evaluation investigates the following five research questions (RQ):

\begin{itemize}[leftmargin=*]
\item \textbf{RQ.1}: Are the distilled rules effective for configuration selection, as measured by their improvement to LLM-based selection? (\Cref{sec:imp_LLMs})

\item \textbf{RQ.2}: How effective is \system compared with bottom-up search methods, and direct-selection methods, including nearest-neighbor retrieval and LLM-based approaches? (\Cref{sec:cmp_base})

\item \textbf{RQ.3}: How well do \system, LLMs, and bottom-up methods adapt to new circuits and devices beyond \bench? (\Cref{sec:scala})

\item \textbf{RQ.4}: Does \system retain its advantage in noise-aware execution fidelity? (\Cref{sec:real_fidelity})

\item \textbf{RQ.5}: How do the key components and parameters affect the performance of \system? (\Cref{sec:aba_study})

\end{itemize}

\subsection{Evaluation Settings}\label{sec:exp_setup}

\noindent
\textbf{LLMs.}
We use the same ten LLMs as in \Cref{sec:eva_settings}.

\noindent
\textbf{Evaluation Metric.}
We use the $F_{eff}$ improvement over $O_3^{DA}$ as the primary metric, together with the $p$-value introduced in \Cref{sec:eva_settings}, and report compilation time. For RQ.1--RQ.3 and RQ.5, the improvement is computed using $F_{\mathrm{eff}}$ defined in \Cref{eq:fidelity}. For RQ.4, we instead use noise-aware execution fidelity obtained from noisy simulation with 2,048 shots per execution.

\noindent
\textbf{Baselines.}
For bottom-up baselines, we evaluate reinforcement learning (RL) and Bayesian optimization (BO). We re-implement MQT Predictor~\cite{mqtpredictor}, a two-stage approach that uses ML for device selection and PPO~\cite{ppo} for compiler-option tuning, under Qiskit V2.4.1 (originally developed for V0.39.2). To retrain both stages under the updated Qiskit version, we construct the training pool using the original train dataset used by MQT Predictor together with QCircuitBench~\cite{yang2025qcircuitbench}, a large-scale quantum-circuit dataset. After removing circuits overlapping with \bench, we sample 2,000 circuits, four times the original training-set size of 500, using 95\% for training and 5\% for validation. Following the MQT Predictor, which trains seven candidate ML models to predict the device that achieves the highest estimated fidelity under the $O_3$ setting and selects the best-performing model for device selection, we train the same seven models. In our setting, Random Forest and Decision Tree achieve the highest validation accuracy and are included in the evaluation.

For option selection, we train Maskable PPO on the same 2,000 circuits, following MQT Predictor by masking capacity-infeasible device choices and using the same hyperparameters. We only increase the training steps from the original 100,000 to 1,000,000 and fix the random seed to 42. The state consists of the seven circuit features, and the reward is based on the achieved $F_{\mathrm{eff}}$. The model is trained on the server described in \Cref{sec:eva_settings}, using one NVIDIA H100 GPU with 80 GB of memory.

We also evaluate BO~\cite{boca,cobyan,PDCAT} over the same compiler-option space, using 10 initial random samples, 50 iterations, and early stopping after 20 consecutive non-improving iterations, following the parameter setting used in prior quantum compiler pass tuning work~\cite{zhong2026qutunerfeaturelearningguidedoptimization}. To isolate the impact of device selection, we further construct \emph{Joint RL/BO}, which jointly optimize the device and compiler options, and \emph{Device-Aware RL/BO}, which fix the capacity-feasible device with the lowest average gate error before option selection, as detailed in \Cref{sec:level_distribution}.

For top-down selection, we include a nearest-neighbor retrieval baseline (NN-Retrieval), which retrieves the most similar circuit among the 68 rule-distillation cases using the seven circuit features in \Cref{sec:tasks} and reuses its configuration. If the retrieved device is infeasible, it is replaced by the capacity-feasible device with the lowest average gate error while retaining the compiler options.

\noindent
\textbf{Prompt Setting.}
The rule-guided prompt extends the prompt in \Cref{sec:eva_settings} with a rule section.

\noindent
\textbf{Benchmarks.}
We use \bench for RQ.1--RQ.2 and RQ.4--RQ.5, evaluating \textbf{only the 200 cases} not used for rule distillation to avoid overfitting. For RQ.3, we additionally sample 100 QCircuitBench circuits~\cite{yang2025qcircuitbench} that overlap with neither \bench nor the 2,000 RL training circuits.

\noindent
\textbf{Implementation Details.}
The LLM inference settings follow \Cref{sec:eva_settings}. For \system, we split the 68 rule-distillation cases into 54 training cases and 14 validation cases (detailed in \Cref{sec:predictive_training}), yielding 51,840 device-option training samples and 13,440 validation samples before filtering infeasible configurations. The hyperparameters $\tau_{\mathrm{seek}}=0.1$, $\tau_{\mathrm{safe}}=0.002$, $\tau_{\mathrm{risk}}=0.6$, $\lambda_r=1$, and $\lambda_t=0.002$ are selected on the 14-case validation set by maximizing the average $F_{eff}$ improvement over $O_3^{DA}$.

\begin{table*}[t]
\centering
\caption{Configuration selection performance of evaluated LLMs across three levels. In the Effective Estimated Fidelity columns, "LLM" denotes the $F_{eff}$ improvement of the LLM-generated configuration over $O_3^{DA}$, while "Oracle" denotes the $F_{eff}$ improvement of the oracle configuration over $O_3^{DA}$.}
\label{tab:llm_imp}
\setstretch{.9}
\begin{adjustbox}{max width=\linewidth}
\setlength{\tabcolsep}{1.ex}

\begin{tabular}{
ll|rrr|rrr|r|rr|rr
}

\toprule

\multirow{3}{*}{Source}
& \multirow{3}{*}{LLMs}
& \multicolumn{7}{c|}{\textbf{Effective Estimated Fidelity ($F_{eff}$ in \Cref{eq:fidelity})}}
& \multicolumn{4}{c}{\textbf{Compilation Time}} \\

\cline{3-9}
\cline{10-13}

&
& \multicolumn{3}{c}{\textbf{-w/o Rules}}
& \multicolumn{4}{c|}{\textbf{-w Rules}}
& \textbf{-w/o R}
& \textbf{-w R}
& \multirow{2}{*}{$O_3^{DA}$}
& \multirow{2}{*}{Oracle} \\

\cline{3-9}
\cline{10-11}

&
& B/E/W $O_3^{DA}$
& LLM
& $p_{\mathrm{LLM}}$
& B/E/W $O_3^{DA}$
& LLM
& $p_{\mathrm{LLM}}$
& Oracle
& LLM
& LLM
& &
\\

\midrule


\multicolumn{13}{c}{\textbf{Level 1: 76 Cases}} \\
\cline{1-13}

\multirow{5}{*}{\shortstack{Closed-\\source}}
& Seed-2.0-Lite
& 1/11/64
& $-51.88\%$
& $<0.001$
& 1/33/42
& $-15.89\%$
& $<0.001$
& \multirow{10}{*}{$0.03\%$}
& 10.76
& 3.17
& \multirow{10}{*}{3.52}
& \multirow{10}{*}{0.65}
\\

& Gemini-3.1-Flash
& 0/24/52
& $-3.64\%$
& $<0.001$
& 2/48/26
& $-1.50\%$
& $<0.001$
&
& 4.87
& 2.91
& &
\\

& Gemini-3.5-Flash
& 0/24/52
& $-3.64\%$
& $<0.001$
& 1/59/16
& $-3.63\%$
& $<0.001$
&
& 4.87
& 0.75
& &
\\

& GPT-5-Mini
& 0/25/51
& $-4.32\%$
& $<0.001$
& 3/52/21
& $-1.64\%$
& $<0.001$
&
& 4.87
& 0.78
& &
\\

& GPT-5.5
& 0/39/37
& $-2.77\%$
& $<0.001$
& 2/59/15
& $\textbf{-1.20\%}$
& $0.001$
&
& 4.47
& 0.84
& &
\\

\cline{1-8}
\cline{10-11}

\multirow{5}{*}{\shortstack{Open-\\source}}
& DeepSeek-3.2
& 0/38/38
& $-2.63\%$
& $<0.001$
& 2/60/14
& $-1.79\%$
& $0.003$
&
& 4.87
& 2.90
& &
\\

& DeepSeek-V4-Flash
& 0/24/52
& $-7.13\%$
& $<0.001$
& 2/58/16
& $-2.16\%$
& $<0.001$
&
& 4.88
& 0.68
& &
\\

& GLM-5
& 0/24/52
& $-8.70\%$
& $<0.001$
& 2/55/19
& $-2.69\%$
& $<0.001$
&
& 10.46
& 0.75
& &
\\

& Qwen3.5-27B
& 0/24/52
& $-3.64\%$
& $<0.001$
& 2/57/17
& $\textbf{-1.44\%}$
& $<0.001$
&
& 4.87
& 3.10
& &
\\

& Qwen3.5-397B-A17B
& 0/26/50
& $-15.13\%$
& $<0.001$
& 1/49/26
& $-3.26\%$
& $<0.001$
&
& 4.87
& 3.11
& &
\\

\midrule


\multicolumn{13}{c}{\textbf{Level 2: 67 Cases}} \\
\cline{1-13}

\multirow{5}{*}{\shortstack{Closed-\\source}}
& Seed-2.0-Lite
& 7/0/60
& $-40.95\%$
& $<0.001$
& 29/6/32
& $-11.31\%$
& $0.136$
& \multirow{10}{*}{$6.04\%$}
& 0.11
& 0.11
& \multirow{10}{*}{0.09}
& \multirow{10}{*}{0.13}
\\

& Gemini-3.1-Flash
& 6/5/56
& $-4.39\%$
& $<0.001$
& 21/31/15
& $0.27\%$
& $0.777$
&
& 0.10
& 0.10
& &
\\

& Gemini-3.5-Flash
& 5/6/56
& $-2.97\%$
& $<0.001$
& 18/37/12
& $0.40\%$
& $0.318$
&
& 0.10
& 0.11
& &
\\

& GPT-5-Mini
& 6/7/54
& $-2.84\%$
& $<0.001$
& 41/7/19
& $\textbf{0.59\%}$
& $0.010$
&
& 0.10
& 0.14
& &
\\

& GPT-5.5
& 3/33/31
& $-3.16\%$
& $<0.001$
& 32/26/9
& $0.46\%$
& $0.007$
&
& 0.10
& 0.11
& &
\\

\cline{1-8}
\cline{10-11}

\multirow{5}{*}{\shortstack{Open-\\source}}
& DeepSeek-3.2
& 2/24/41
& $-3.84\%$
& $<0.001$
& 29/20/18
& $-1.66\%$
& $0.262$
&
& 9.05
& 0.10
& &
\\

& DeepSeek-V4-Flash
& 6/5/56
& $-3.73\%$
& $<0.001$
& 36/18/13
& $0.78\%$
& $0.003$
&
& 0.10
& 0.12
& &
\\

& GLM-5
& 6/4/57
& $-4.13\%$
& $<0.001$
& 30/22/15
& $-0.22\%$
& $0.081$
&
& 0.10
& 0.11
& &
\\

& Qwen3.5-27B
& 6/5/56
& $-3.45\%$
& $<0.001$
& 37/16/14
& $-0.31\%$
& $0.022$
&
& 0.10
& 0.14
& &
\\

& Qwen3.5-397B-A17B
& 7/5/55
& $-3.41\%$
& $<0.001$
& 42/11/14
& $\textbf{0.99\%}$
& $0.002$
&
& 0.10
& 0.12
& &
\\

\midrule


\multicolumn{13}{c}{\textbf{Level 3: 57 Cases}} \\
\cline{1-13}

\multirow{5}{*}{\shortstack{Closed-\\source}}
& Seed-2.0-Lite
& 13/13/31
& $-16.96\%$
& $0.001$
& 30/6/21
& $7.46\%$
& $0.265$
& \multirow{10}{*}{$31.92\%$}
& 0.18
& 0.22
& \multirow{10}{*}{0.19}
& \multirow{10}{*}{0.26}
\\

& Gemini-3.1-Flash
& 2/30/25
& $-0.59\%$
& $<0.001$
& 20/33/4
& $11.72\%$
& $0.002$
&
& 0.18
& 0.19
& &
\\

& Gemini-3.5-Flash
& 2/30/25
& $-0.59\%$
& $<0.001$
& 23/21/13
& $16.03\%$
& $0.044$
&
& 0.18
& 0.24
& &
\\

& GPT-5-Mini
& 2/31/24
& $-0.59\%$
& $0.001$
& 36/7/14
& $13.88\%$
& $0.005$
&
& 0.18
& 0.23
& &
\\

& GPT-5.5
& 3/39/15
& $-0.10\%$
& $0.112$
& 35/15/7
& $\textbf{18.87\%}$
& $<0.001$
&
& 0.18
& 0.23
& &
\\

\cline{1-8}
\cline{10-11}

\multirow{5}{*}{\shortstack{Open-\\source}}
& DeepSeek-3.2
& 1/40/16
& $-0.30\%$
& $0.005$
& 15/16/26
& $12.02\%$
& $0.683$
&
& 0.19
& 0.18
& &
\\

& DeepSeek-V4-Flash
& 3/26/28
& $-0.72\%$
& $<0.001$
& 30/14/13
& $14.96\%$
& $0.005$
&
& 0.18
& 0.21
& &
\\

& GLM-5
& 3/28/26
& $-3.39\%$
& $<0.001$
& 20/24/13
& $10.40\%$
& $0.221$
&
& 0.18
& 0.21
& &
\\

& Qwen3.5-27B
& 2/30/25
& $-0.59\%$
& $<0.001$
& 32/10/15
& $13.71\%$
& $0.011$
&
& 0.18
& 0.24
& &
\\

& Qwen3.5-397B-A17B
& 3/30/24
& $-0.57\%$
& $0.003$
& 39/6/12
& $\textbf{20.37\%}$
& $0.001$
&
& 0.18
& 0.24
& &
\\

\bottomrule

\end{tabular}
\end{adjustbox}
\vspace{-2.ex}
\end{table*}

\subsection{Effectiveness of Distilled Rules for Configuration Selection (RQ.1)}\label{sec:imp_LLMs}
As shown in \Cref{tab:llm_imp}, incorporating the distilled rules consistently improves LLM performance on quantum compiler configuration selection across all LLMs and all three levels, providing direct evidence of the effectiveness of the distilled rules. For Levels 2 and 3, most LLMs achieve positive gains over $O_3^{DA}$; notably, Qwen3.5-397B-A17B improves over $O_3^{DA}$ by 20.37\% on average for Level 3. For Level 1, however, most models still show slight degradation, with Seed-2.0-Lite exhibiting a larger drop of 15.89\%. These results show that the distilled rules help LLMs identify better configurations when sufficient optimization headroom exists. However, LLMs still struggle on Level 1, where $O_3^{DA}$ is already near optimal and unnecessary changes should be avoided. The statistical results show the same trend: all rule-guided LLMs remain significantly different from $O_3^{DA}$ on Level 1, whereas on Level 3, seven of ten models achieve statistically significant positive gains ($p<0.05$); Level 2 shows a more mixed pattern.

\begin{table*}[t]
\centering
\caption{Comparison of different methods for configuration selection across three difficulty levels. Compile (s) and Select (s) denote the time of compilation and selection.}
\label{tab:cmp_baselines}
\begin{adjustbox}{max width=\linewidth}
\setstretch{.9}
\setlength{\tabcolsep}{3.ex}
\begin{tabular}{llccccc}
\hline
\textbf{Paradigm}
& \textbf{Method}
& \textbf{B/E/W $O_3^{DA}$}
& \textbf{$F_{eff}$ Imp. over $O_3^{DA}$}
& \textbf{$p$}
& \textbf{Compile (s)}
& \textbf{Select (s)} \\
\hline

\multicolumn{7}{c}{\textbf{Level 1}} \\
\hline

\multirow{8}{*}{Bottom-up}
& Random Forest + BO & 1/24/51 & $-32.65\%$ & $<0.001$ & 11.09 & 3320.86 \\
& Decision Tree + BO & 2/30/44 & $-23.72\%$ & $<0.001$ & 1.78 & 2944.71 \\
& Joint BO & 4/41/31 & $-8.27\%$ & $<0.001$ & 19.01 & 1746.02 \\
& Device-Aware BO & 3/73/0 & $\textbf{0.014\%}$ & 0.109  & 0.81 & 2856.87 \\
\cline{2-7}
& Random Forest + RL & 0/14/62 & $-30.90\%$ & $<0.001$ & 5.22 & 0.15 \\
& Decision Tree + RL & 0/14/62 & $-36.53\%$ & $<0.001$ & 4.11 & 0.15 \\
& Joint RL & 0/21/55 & $-17.31\%$ & $<0.001$ & 4.81 & 0.16 \\
& Device-Aware RL & 0/21/55 & $-19.43\%$ & $<0.001$ & 2.39 & 0.15 \\
\hline

\multirow{4}{*}{Top-down}
& NN-Retrieval & 1/32/43 & $-17.61\%$ & $<0.001$ & 3.61 & 0.001 \\
& GPT-5.5 (-w Rules) & 2/59/15 & $-1.20\%$ & 0.001 & 0.84 & 17.43 \\
& Qwen3.5-397B (-w Rules) & 1/49/26 & $-3.26\%$ & $<0.001$ & 3.11 & 110.73 \\
\cline{2-7}
& \system & 4/67/5 & $0.002\%$ & 0.859 & 1.09 & 3.28 \\
\hline

\multicolumn{7}{c}{\textbf{Level 2}} \\
\hline

\multirow{8}{*}{Bottom-up}
& Random Forest + BO & 29/6/32 & $-17.45\%$ & 0.013 & 0.18 & 2462.41 \\
& Decision Tree + BO & 28/6/33 & $-19.99\%$ &  0.007 & 0.21 & 2512.41 \\
& Joint BO & 38/10/19 & $-5.25\%$ & 0.439 & 9.14 & 820.02 \\
& Device-Aware BO & 43/24/0 & $2.45\%$ & $<0.001$ & 0.11 & 2146.68 \\
\cline{2-7}
& Random Forest + RL & 13/3/51 & $-21.91\%$ & $<0.001$ & 0.21 & 0.01 \\
& Decision Tree + RL & 10/3/54 & $-30.53\%$ & $<0.001$ & 0.20 & 0.01 \\
& Joint RL & 17/2/48 & $-8.72\%$ & $<0.001$ & 0.13 & 0.01 \\
& Device-Aware RL & 14/10/43 & $-13.39\%$ & $<0.001$ & 0.18 & 0.01 \\
\hline

\multirow{4}{*}{Top-down}
& NN-Retrieval & 19/22/26 & $-16.14\%$ & 0.005 & 0.12 & 0.001 \\
& GPT-5.5 (-w Rules) & 32/26/9 & $0.46\%$ & 0.007 & 0.11 & 17.39 \\
& Qwen3.5-397B (-w Rules) & 42/11/14 & $0.99\%$ & 0.002 & 0.12 & 110.98 \\
\cline{2-7}
& \system & 42/20/5 & $\textbf{3.47\%}$ & $<0.001$ & 0.12 & 3.30 \\
\hline

\multicolumn{7}{c}{\textbf{Level 3}} \\
\hline

\multirow{8}{*}{Bottom-up}
& Random Forest + BO & 44/2/11 & $13.65\%$ & $<0.001$ & 53.41 & 1535.58 \\
& Decision Tree + BO & 33/1/23 & $17.87\%$ & 0.007 & 21.38 & 1681.89 \\
& Joint BO & 47/9/1 & $21.54\%$ & $<0.001$ & 0.28 & 967.80 \\
& Device-Aware BO & 38/19/0 & $10.62\%$ & $<0.001$ & 0.21 & 1504.36 \\
\cline{2-7}
& Random Forest + RL & 37/1/19 & $21.26\%$ & 0.002 & 0.35 & 0.02 \\
& Decision Tree + RL & 31/1/25 & $15.66\%$ & 0.106 & 0.36 & 0.02 \\
& Joint RL & 39/1/17 & $24.49\%$ & $<0.001$ & 0.31 & 0.02 \\
& Device-Aware RL & 38/3/16 & $18.24\%$ & 0.009 & 0.30 & 0.02 \\
\hline

\multirow{4}{*}{Top-down}
& NN-Retrieval & 28/15/14 & $5.24\%$ & 0.114 & 0.24 & 0.001 \\
& GPT-5.5 (-w Rules) & 35/15/7 & $18.87\%$ & $<0.001$ & 0.23 & 17.06 \\
& Qwen3.5-397B (-w Rules) & 39/6/12 & $20.37\%$ & 0.001 & 0.24 & 108.54 \\
\cline{2-7}
& \system & 49/7/1 & $\textbf{28.56\%}$ & $<0.001$ & 0.25 & 3.33 \\
\hline

\end{tabular}
\end{adjustbox}
\vspace{-1.ex}
\end{table*}

\begin{figure*}[t]
    \centering
    \begin{minipage}[t]{0.48\linewidth}
        \centering
        \includegraphics[width=\linewidth]{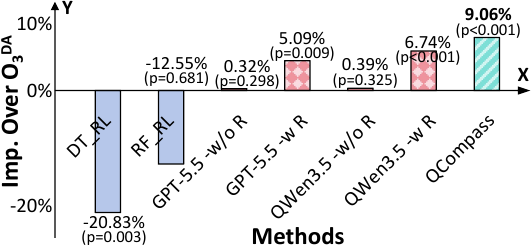}
        \caption{Scalable analysis of \system and baselines on new circuits and devices.}
        \label{fig:cmp_scalability}
    \end{minipage}
    \hfill
    \begin{minipage}[t]{0.48\linewidth}
        \centering
        \includegraphics[width=\linewidth]{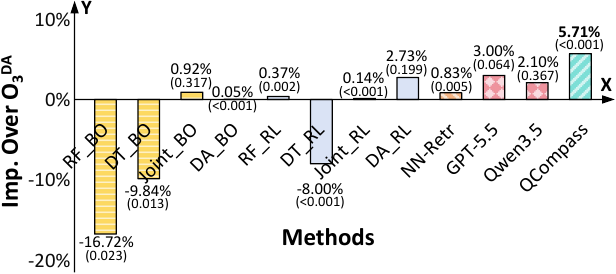}
        \caption{Comparison of noise-aware execution fidelity on QASMBench small and medium cases.}
        \label{fig:cmp_real}
    \end{minipage}
    \Description{xx}
\end{figure*}

\subsection{Effectiveness of \system (RQ.2)}\label{sec:cmp_base}

We further compare \system with two bottom-up methods (RL and BO), the NN-Retrieval method, the two best-performing LLMs in \Cref{tab:llm_imp}, GPT-5.5 and Qwen3.5-397B-A17B. As shown in \Cref{tab:cmp_baselines}, \system achieves positive gains over $O_3^{DA}$ across all three levels and outperforms all baselines on Levels 2 and 3, demonstrating its effectiveness. On Level 1, \system achieves a slightly lower gain than Device-Aware BO, with a difference of only 0.012 percentage points. However, \system requires only 3.28 s search time per case on average, compared with 2,856.87 s for Device-Aware BO. 

Notably, on Level 1, \system \emph{matches} $O_3^{DA}$ on 67 of 76 cases, with nearly balanced better and worse cases (4 vs. 5), resulting in no statistically significant difference ($p=0.859$). By comparison, the Device-aware BO also largely matches $O_3^{DA}$ (73 of 76 cases) but improves all three remaining cases, leading to a lower yet still non-significant $p$-value ($p=0.109$). In contrast, \system's improvements on Levels 2 and 3 are statistically significant ($p<0.001$), consistent with its goal of \textbf{preserving near-optimal configurations while exploiting larger optimization headroom}.

For BO and RL, most variants underperform $O_3^{DA}$ on Levels 1 and 2, except Device-Aware BO. This suggests that device-selection errors can trap BO on suboptimal devices, while RL tends to make unnecessary option changes even when the $O_3^{DA}$ device is fixed. On Level 3, however, most BO/RL variants outperform Device-Aware BO, indicating that larger optimization headroom allows alternative device choices and more aggressive option tuning to provide greater gains.
NN-Retrieval underperforms Device-Aware RL/BO  across all three levels, achieving only a 5.24\% improvement over $O_3^{DA}$ on Level 3. This indicates that directly reusing configurations from similar circuits is insufficient, while distillation of transferable rules is considerably more effective.

From the perspective of search time, \system requires only about 3.30 seconds per case on average, substantially faster than GPT-5.5 (about 17 seconds), Qwen3.5-397B-A17B (about 110 seconds), and BO (820--3,320 seconds). Although \system is slower than RL and NN-Retrieval, which require only about 0.001--0.15 seconds per case, it achieves substantially better configuration quality, making this additional inference cost a practical trade-off.

\subsection{Adaptation to New Devices and Circuits (RQ.3)}\label{sec:scala}

We further evaluate the adaptation ability of the distilled rules and \system on new circuits and devices. Specifically, we randomly sample 100 circuits from QCircuitBench that do not overlap with \bench and the RL training dataset, and replace the original device pool with four new devices: \texttt{FakeTorino} (133 qubits), \texttt{FakeSherbrooke} (127 qubits), \texttt{FakePrague} (33 qubits), and \linebreak \texttt{FakeMontrealV2} (27 qubits), and all methods are restricted to selecting among them. For the ML-based device selectors, we follow their extensible design and retrain them by expanding the training device pool to include the four new devices, while restricting evaluation selection to the four new devices. The RL option-selection policies and \system are applied directly without retraining.

As shown in \Cref{fig:cmp_scalability}, \system achieves a 9.06\% average improvement over $O_3^{DA}$, while both rule-guided LLMs achieve gains above 5\%. In contrast, the two ML+RL methods yield negative improvements. These results indicate that \system's top-down selection generalizes more effectively to new circuits and devices than the bottom-up RL-based approaches.

\subsection{Evaluation with Noise-Aware Execution Fidelity on the Noisy Simulator (RQ.4)}\label{sec:real_fidelity}

In \Cref{eq:fidelity}, we use estimated fidelity to evaluate compiled circuits based on device error information without actual execution. To verify whether our conclusions remain consistent under noise-aware execution fidelity, we conduct an additional evaluation in this section. Specifically, we compute the Hellinger fidelity~\cite{fidelity} as our evaluation metric. For each selected device, we construct an \texttt{Aer} noise model using the noise model from the corresponding fake backend, which incorporates backend-specific error parameters.
All methods in \Cref{tab:cmp_baselines} are included for evaluation. Since noise-aware execution fidelity simulation for large or complex circuits incurs substantial memory and execution costs, we evaluate 63 small and medium circuits from QASMBench, as detailed in \Cref{table:base_bench}.

As shown in \Cref{fig:cmp_real}, \system still achieves the largest mean improvement, while both rule-guided LLMs also maintain positive gains. On the evaluated set, \system achieves a 5.71\% mean improvement over \(O_3^{DA}\), with the statistical significance \(p<0.001\). Therefore, the noise-aware evaluation remains directionally consistent with our estimated-fidelity results, although the magnitude of the improvement differs from that in \Cref{tab:cmp_baselines}.

\subsection{Ablation and Sensitivity Study (RQ.5)}\label{sec:aba_study}

\begin{table}
    \centering
    \caption{Ablation study of the \system.}
    \label{tab:qutop_ablation}
    \setstretch{.9}
    \setlength{\tabcolsep}{4.ex}
    \resizebox{\linewidth}{!}{
    \begin{tabular}{lccc}
        \toprule
        \textbf{Method} & \textbf{Level 1} & \textbf{Level 2} & \textbf{Level 3} \\
        \midrule

        \multicolumn{4}{c}{\textbf{Components}} \\
        \midrule

        \system
        & $0.002\%$
        & $3.47\%$
        & $28.56\%$ \\

        \multicolumn{1}{r}{-w/o Policy}
        & $-13.38\%$
        & $-3.21\%$
        & $17.36\%$ \\

        \multicolumn{1}{r}{-w/o Risk}
        & $-27.13\%$
        & $-43.06\%$
        & $15.13\%$ \\

        \multicolumn{1}{r}{-w/o Rules}
        & $-0.55\%$
        & $-28.72\%$
        & $19.69\%$ \\

        \midrule
        \multicolumn{4}{c}{\textbf{Distilling Model}} \\
        \midrule

        \multicolumn{1}{r}{Claude-4.6}
        & $-1.81\%$
        & $-3.51\%$
        & $20.71\%$ \\

        \multicolumn{1}{r}{GPT-5.2}
        & $-14.10\%$
        & $-2.40\%$
        & $18.66\%$ \\

        \bottomrule
    \end{tabular}
    }
\end{table}
We further analyze \system's key components, the choice of distilling LLM, and its key parameters. For component ablation, \emph{-w/o Policy} removes the policy gate, so all cases undergo candidate generation and scoring without distinguishing whether $O_3^{DA}$ should be preserved or only locally modified; \emph{-w/o Risk} removes the timeout risk-aware safeguard; and \emph{-w/o Rules} replaces rule-guided candidate generation with the full feasible device-option space. 

As shown in \Cref{tab:qutop_ablation}, removing the policy gate causes the largest degradation on Level 1, confirming the importance of avoiding unnecessary exploration when $O_3^{DA}$ is already near optimal. Removing Risk also substantially degrades Levels 1 and 2 because some selected configurations time out, yielding extremely large negative improvements over $O_3^{DA}$ that dominate the average. Level 3 remains positive because its larger optimization headroom allows successful selections to offset part of these losses. In contrast, removing the distilled rules mainly hurts Levels 2 and 3, especially Level 2, showing that the rules are most useful when meaningful optimization headroom exists by guiding the search toward promising configuration regions.

\begin{figure*}[ht]
    \centering
    \includegraphics[width=\linewidth]{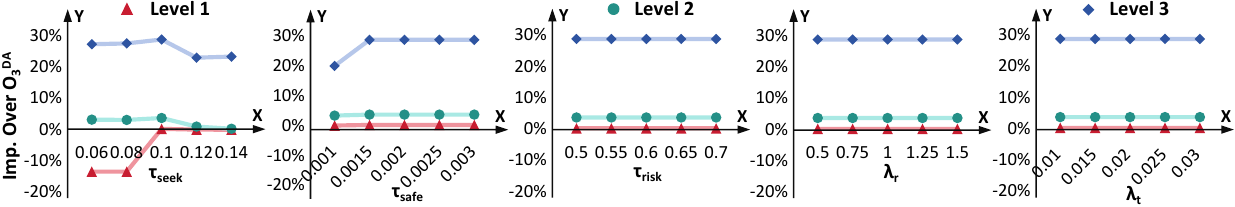}
    \caption{The sensitivity analysis of the \system.}
    \Description{xx}
    \label{fig:sensitivity}
    \vspace{-.ex}
\end{figure*}

For the distilling-LLM ablation, we replace Claude-Opus-5 with two alternative LLMs: Claude-Sonnet-4.6~\cite{Web:claude46}, an earlier model from the same Claude family, and GPT-5.2~\cite{Web:gpt52}, from a different model family. We repeat the rule-distillation process described in \Cref{sec:rule_gen}. Claude-Sonnet-4.6 generates 17 rules, whereas GPT-5.2 generates only 7 rules.

\Cref{tab:qutop_ablation} shows that replacing the Claude-Opus-5 distilled rules consistently weakens \system. Claude-Sonnet-4.6 produces even more rules than Opus-5 (17 vs. 13), yet its gains drop to $-1.81\%$, $-3.51\%$, and $20.71\%$ across Levels 1--3, indicating that rule quantity alone is insufficient. Its rules rely more on rigid thresholds and can conflict; for example, one rule favors $O_3$ with SABRE for wide, highly entangled circuits, while another downgrades high two-qubit-error-exposure (often associated with heavily entangling circuits) circuits to $O_0$--$O_1$ with trivial layout. GPT-5.2 exhibits a different limitation, distilling only seven rules, which provides weaker coverage of the configuration space and leads to $-14.10\%$, $-2.40\%$, and $18.66\%$. In contrast, the Claude-Opus-5 distilled rules more systematically capture interactions among circuit structure, device characteristics, and optimization opportunity, yielding more transferable guidance.

For ablation groups in \Cref{tab:qutop_ablation}, paired Wilcoxon signed-rank tests between \system and each ablated variant further confirm that the observed degradations are statistically significant (\(p<0.05\)).

For parameter sensitivity, we vary $\tau_{\mathrm{seek}}$, $\tau_{\mathrm{safe}}$, $\tau_{\mathrm{risk}}$, $\lambda_r$, and $\lambda_t$ around their default settings in \Cref{sec:exp_setup}, as shown in \Cref{fig:sensitivity}. Among them, $\tau_{\mathrm{seek}}$ has the largest impact: the default value performs best, while either increasing or decreasing it degrades performance. For $\tau_{\mathrm{safe}}$, setting it to 0.001 reduces Level-3 performance, whereas values above 0.0015 produce nearly unchanged results across all three levels. In contrast, $\tau_{\mathrm{risk}}$, $\lambda_r$, and $\lambda_t$ are relatively robust, with no noticeable performance variation within the evaluated ranges.

\section{Related Work}

\noindent
\textbf{Compiler Auto-tuning.}
Classical compiler auto-tuning has explored heuristic search~\cite{boca,cobyan,PDCAT,gene1,gene2,gene3}, RL-based pass selection~\cite{rl1,rl2,rl3,rl4,oopsla_phase,autophase}, similarity-based retrieval~\cite{ICS_Similarity,SC_similarity}, and LLM-guided optimization~\cite{pan2026eccoevidencedrivencausalreasoning,qiu2026passbypassoptimizationintentdrivenir,meta_llm_compiler}. Quantum compiler tuning has also adopted heuristic and RL-based methods~\cite{mqtpredictor,tuniq,DAC23,quantum_tuning_cibda,mills2026reinforcementlearningadaptivecomposition,dangwal2025cliffordassistedoptimalpass}. However, only two of these studies~\cite{DAC23,mqtpredictor} focus on compiler configuration selection with the bottom-up RL method. In this paper, we explore a top-down selection paradigm, achieving better performance than existing methods.

\noindent
\textbf{Quantum Software Engineering}.
Recent advances in quantum computing have led to increasing interest in quantum software engineering~\cite{qse_roadmap}, encompassing areas such as quantum compiler tuning~\cite{DAC23,mills2026reinforcementlearningadaptivecomposition,dangwal2025cliffordassistedoptimalpass,quantum_tuning_cibda}, execution performance and time prediction~\cite{tosem_circuits,quest}, testing~\cite{qse_test1,qse_test2,qse_test3,qse_test4}, program repair~\cite{qse_repair}, and quantum code generation~\cite{qse_codegen1,qse_codegen2,ASE_Quanbench}.

\section{Threats to Validity}\label{sec:threats_to_valid}

\textbf{Internal Validity}. 
The main internal threat lies in oracle construction, which is bounded by the selected devices, option space, timeout policy, and estimated fidelity proxy. We mitigate this by using identical devices, option spaces and settings for all methods. Another threat comes from our re-implementation of MQT Predictor~\cite{mqtpredictor}. Since the original MQT Predictor was developed under Qiskit V0.39.2, its search space is not identical to our Qiskit V2.4.1 setting, and our reproduced models therefore cannot exactly match the original implementation. To mitigate this threat, we follow its reported two-stage methodology and model choices and adopt the same training hyperparameters, except that we increase the training steps by 10$\times$. Finally, LLM nondeterminism and parsing errors may affect results; we mitigate them by fixing the temperature to 0.0 and seed to 42. Another internal threat arises from the manual verification of the LLM-distilled rules and may therefore introduce researcher judgment. We mitigate this threat by checking each distilled rule against its supporting cases and removing redundant or overly case-specific statements.

\textbf{External Validity}. 
Our results are based on the evaluated circuit benchmarks, fake backends, Qiskit V2.4.1, and selected compiler options. Although \bench contains diverse QASMBench and VeriQBench circuits, the distilled rules and \system may not fully generalize to substantially different workloads, hardware architectures, or future compiler versions. Moreover, most experiments use estimated fidelity, while real devices may additionally be affected by calibration drift, crosstalk, and runtime noise. We partially mitigate these threats through evaluations on new circuits and devices (\Cref{sec:scala}) and an additional noise-aware execution fidelity experiment on the noisy simulator (\Cref{sec:real_fidelity}), but broader compiler and real-hardware evaluations remain future work.

\section{Conclusion}

In this paper, we explore a \emph{top-down} paradigm for quantum compiler configuration selection. We construct \bench with empirical oracle configurations, use these oracles to distill transferable configuration rules, and apply them to new circuits. We further develop \system, a rule-guided framework for fast configuration selection. Our results show that \system achieves best overall quality–latency trade-off compared with baselines.

In future work, we plan to extend this top-down paradigm to broader quantum compilers, larger and evolving device pools, and real quantum hardware. We also aim to explore more efficient oracle construction to reduce offline characterization cost and improve scalability, together with continual rule adaptation to preserve the effectiveness and transferability of distilled rules.

\bibliographystyle{ACM-Reference-Format}
\bibliography{bibfile}

@String{Computing = "Computing" }

@String{Computer = "{IEEE} Computer" }

@String{Academic = "Academic Press" }

@inproceedings{meta_llm_compiler,
author = {Cummins, Chris and Seeker, Volker and Grubisic, Dejan and Roziere, Baptiste and Gehring, Jonas and Synnaeve, Gabriel and Leather, Hugh},
title = {LLM Compiler: Foundation Language Models for Compiler Optimization},
year = {2025},
isbn = {9798400714078},
publisher = {Association for Computing Machinery},
address = {New York, NY, USA},
url = {https://doi.org/10.1145/3708493.3712691},
doi = {10.1145/3708493.3712691},
booktitle = {Proceedings of the 34th ACM SIGPLAN International Conference on Compiler Construction},
pages = {141–153},
numpages = {13},
location = {Las Vegas, NV, USA},
series = {CC '25}
}

@article{feng2026llmrouter,
  title={LLMRouter: Unified Infrastructure for Developing, Evaluating, and Deploying LLM Routers},
  author={Feng, Tao and Yu, Fangxu and Zhang, Haozhen and Dai, Zhongjie and Yuan, Liangqi and Lei, Zijie and Zhang, Weizhi and Zhu, Kunlun and Yue, Haodong and Xuan, Keyang and Liu, Ge and You, Jiaxuan},
  journal={arXiv preprint arXiv:2608.06867},
  year={2026}
}

@inproceedings{boca,
author = {Chen, Junjie and Xu, Ningxin and Chen, Peiqi and Zhang, Hongyu},
title = {Efficient Compiler Autotuning via Bayesian Optimization},
year = {2021},
isbn = {9781450390859},
publisher = {IEEE Press},
url = {https://doi.org/10.1109/ICSE43902.2021.00110},
doi = {10.1109/ICSE43902.2021.00110},
booktitle = {Proceedings of the 43rd International Conference on Software Engineering},
pages = {1198–1209},
numpages = {12},
location = {Madrid, Spain},
series = {ICSE '21}
}

@inproceedings{ln_fidelity,
author = {Tan, Bochen and Cong, Jason},
title = {Optimal layout synthesis for quantum computing},
year = {2020},
isbn = {9781450380263},
publisher = {Association for Computing Machinery},
address = {New York, NY, USA},
url = {https://doi.org/10.1145/3400302.3415620},
doi = {10.1145/3400302.3415620},
booktitle = {Proceedings of the 39th International Conference on Computer-Aided Design},
articleno = {137},
numpages = {9},
location = {Virtual Event, USA},
series = {ICCAD '20}
}

@misc{qiao2026warmtunerprogramspecificwarmstarts,
      title={WarmTuner: Program-Specific Warm Starts for Compiler Autotuning via Offline-to-Online Reinforcement Learning}, 
      author={Tianlu Qiao and Mingxuan Zhu and Zeyu Sun and Dan Hao},
      year={2026},
      eprint={2607.25831},
      archivePrefix={arXiv},
      primaryClass={cs.SE},
      url={https://arxiv.org/abs/2607.25831}, 
}

@article{PDCAT,
author = {Zhu, Mingxuan and Sun, Zeyu and Hao, Dan},
title = {PDCAT: Preference-Driven Compiler Auto-tuning},
year = {2025},
issue_date = {July 2025},
publisher = {Association for Computing Machinery},
address = {New York, NY, USA},
volume = {2},
number = {FSE},
url = {https://doi.org/10.1145/3715756},
doi = {10.1145/3715756},
journal = {Proc. ACM Softw. Eng.},
month = jun,
articleno = {FSE039},
numpages = {21}
}

@article{cobyan,
author = {Zhu, Mingxuan and Hao, Dan and Chen, Junjie},
title = {Compiler Autotuning through Multiple-phase Learning},
year = {2024},
issue_date = {May 2024},
publisher = {Association for Computing Machinery},
address = {New York, NY, USA},
volume = {33},
number = {4},
issn = {1049-331X},
url = {https://doi.org/10.1145/3640330},
doi = {10.1145/3640330},
journal = {ACM Trans. Softw. Eng. Methodol.},
month = apr,
articleno = {100},
numpages = {38}
}

@inproceedings{gene1,
author = {Garciarena, Unai and Santana, Roberto},
title = {Evolutionary Optimization of Compiler Flag Selection by Learning and Exploiting Flags Interactions},
year = {2016},
isbn = {9781450343237},
publisher = {Association for Computing Machinery},
address = {New York, NY, USA},
url = {https://doi.org/10.1145/2908961.2931696},
doi = {10.1145/2908961.2931696},
booktitle = {Proceedings of the 2016 on Genetic and Evolutionary Computation Conference Companion},
pages = {1159–1166},
numpages = {8},
location = {Denver, Colorado, USA},
series = {GECCO '16 Companion}
}

@inproceedings{gene2,
author = {Pan, Haolin and Wei, Yuanyu and Xing, Mingjie and Wu, Yanjun and Zhao, Chen},
title = {Towards Efficient Compiler Auto-tuning: Leveraging Synergistic Search Spaces},
year = {2025},
isbn = {9798400712753},
publisher = {Association for Computing Machinery},
address = {New York, NY, USA},
url = {https://doi.org/10.1145/3696443.3708961},
doi = {10.1145/3696443.3708961},
booktitle = {Proceedings of the 23rd ACM/IEEE International Symposium on Code Generation and Optimization},
pages = {614–627},
numpages = {14},
location = {Las Vegas, NV, USA},
series = {CGO '25}
}

@inproceedings{gene3,
author = {Hoste, Kenneth and Eeckhout, Lieven},
title = {Cole: compiler optimization level exploration},
year = {2008},
isbn = {9781595939784},
publisher = {Association for Computing Machinery},
address = {New York, NY, USA},
url = {https://doi.org/10.1145/1356058.1356080},
doi = {10.1145/1356058.1356080},
booktitle = {Proceedings of the 6th Annual IEEE/ACM International Symposium on Code Generation and Optimization},
pages = {165–174},
numpages = {10},
location = {Boston, MA, USA},
series = {CGO '08}
}

@inproceedings{rl1,
author = {Coons, Katherine E. and Robatmili, Behnam and Taylor, Matthew E. and Maher, Bertrand A. and Burger, Doug and McKinley, Kathryn S.},
title = {Feature selection and policy optimization for distributed instruction placement using reinforcement learning},
year = {2008},
isbn = {9781605582825},
publisher = {Association for Computing Machinery},
address = {New York, NY, USA},
url = {https://doi.org/10.1145/1454115.1454122},
doi = {10.1145/1454115.1454122},
booktitle = {Proceedings of the 17th International Conference on Parallel Architectures and Compilation Techniques},
pages = {32–42},
numpages = {11},
location = {Toronto, Ontario, Canada},
series = {PACT '08}
}

@inproceedings{rl2,
author = {Mirhoseini, Azalia and Pham, Hieu and Le, Quoc V. and Steiner, Benoit and Larsen, Rasmus and Zhou, Yuefeng and Kumar, Naveen and Norouzi, Mohammad and Bengio, Samy and Dean, Jeff},
title = {Device placement optimization with reinforcement learning},
year = {2017},
publisher = {JMLR.org},
booktitle = {Proceedings of the 34th International Conference on Machine Learning - Volume 70},
pages = {2430–2439},
numpages = {10},
location = {Sydney, NSW, Australia},
series = {ICML'17}
}

@inproceedings{rl3,
author = {Cummins, Chris and Wasti, Bram and Guo, Jiadong and Cui, Brandon and Ansel, Jason and Gomez, Sahir and Jain, Somya and Liu, Jia and Teytaud, Olivier and Steiner, Benoit and Tian, Yuandong and Leather, Hugh},
title = {CompilerGym: robust, performant compiler optimization environments for AI research},
year = {2022},
isbn = {9781665405843},
publisher = {IEEE Press},
url = {https://doi.org/10.1109/CGO53902.2022.9741258},
doi = {10.1109/CGO53902.2022.9741258},
booktitle = {Proceedings of the 20th IEEE/ACM International Symposium on Code Generation and Optimization},
pages = {92–105},
numpages = {14},
location = {Virtual Event, Republic of Korea},
series = {CGO '22}
}

@inproceedings{rl4,
author = {Park, Sunghyun and Latifi, Salar and Park, Yongjun and Behroozi, Armand and Jeon, Byungsoo and Mahlke, Scott},
title = {SRTuner: effective compiler optimization customization by exposing synergistic relations},
year = {2022},
isbn = {9781665405843},
publisher = {IEEE Press},
url = {https://doi.org/10.1109/CGO53902.2022.9741263},
doi = {10.1109/CGO53902.2022.9741263},
booktitle = {Proceedings of the 20th IEEE/ACM International Symposium on Code Generation and Optimization},
pages = {118–130},
numpages = {13},
location = {Virtual Event, Republic of Korea},
series = {CGO '22}
}

@misc{pan2026eccoevidencedrivencausalreasoning,
      title={ECCO: Evidence-Driven Causal Reasoning for Compiler Optimization}, 
      author={Haolin Pan and Lianghong Huang and Jinyuan Dong and Mingjie Xing and Yanjun Wu},
      year={2026},
      eprint={2602.00087},
      archivePrefix={arXiv},
      primaryClass={cs.LG},
      url={https://arxiv.org/abs/2602.00087}, 
}

@inproceedings{
yang2025qcircuitbench,
title={{QC}ircuitBench: A Large-Scale Dataset for Benchmarking Quantum Algorithm Design},
author={Rui Yang and Ziruo Wang and Yuntian Gu and Yitao Liang and Tongyang Li},
booktitle={The Thirty-ninth Annual Conference on Neural Information Processing Systems Datasets and Benchmarks Track},
year={2025},
url={https://openreview.net/forum?id=NkiLldW2bi}
}

@misc{dangwal2025cliffordassistedoptimalpass,
      title={Clifford Assisted Optimal Pass Selection for Quantum Transpilation}, 
      author={Siddharth Dangwal and Gokul Subramanian Ravi and Lennart Maximilian Seifert and Poulami Das and James Sud and Frederic T. Chong},
      year={2025},
      eprint={2306.15020},
      archivePrefix={arXiv},
      primaryClass={quant-ph},
      url={https://arxiv.org/abs/2306.15020}, 
}

@article{mqtpredictor,
author = {Quetschlich, Nils and Burgholzer, Lukas and Wille, Robert},
title = {MQT Predictor: Automatic Device Selection with Device-Specific Circuit Compilation for Quantum Computing},
year = {2025},
issue_date = {March 2025},
publisher = {Association for Computing Machinery},
address = {New York, NY, USA},
volume = {6},
number = {1},
url = {https://doi.org/10.1145/3673241},
doi = {10.1145/3673241},
journal = {ACM Transactions on Quantum Computing},
month = jan,
articleno = {10},
numpages = {26}
}

@article{ppo,
  author       = {John Schulman and
                  Filip Wolski and
                  Prafulla Dhariwal and
                  Alec Radford and
                  Oleg Klimov},
  title        = {Proximal Policy Optimization Algorithms},
  journal      = {CoRR},
  volume       = {abs/1707.06347},
  year         = {2017},
  url          = {http://arxiv.org/abs/1707.06347},
  eprinttype   = {arXiv},
  eprint       = {1707.06347},
  bibsource    = {dblp computer science bibliography, https://dblp.org}
}

@article{Jozsa01121994,
author = {Richard Jozsa},
title = {Fidelity for Mixed Quantum States},
journal = {Journal of Modern Optics},
volume = {41},
number = {12},
pages = {2315--2323},
year = {1994},
publisher = {Taylor \& Francis},
doi = {10.1080/09500349414552171},


URL = { 
    
        https://doi.org/10.1080/09500349414552171
    
    

},
eprint = { 
    
        https://doi.org/10.1080/09500349414552171
    
    

}

}

@inproceedings{DAC23,
author = {Quetschlich, Nils and Burgholzer, Lukas and Wille, Robert},
title = {Compiler Optimization for Quantum Computing Using Reinforcement Learning},
year = {2023},
isbn = {9798350323481},
publisher = {IEEE Press},
url = {https://doi.org/10.1109/DAC56929.2023.10248002},
doi = {10.1109/DAC56929.2023.10248002},
booktitle = {Proceedings of the 60th Annual ACM/IEEE Design Automation Conference},
pages = {1–6},
numpages = {6},
location = {San Francisco, California, United States},
series = {DAC '23}
}

@misc{mills2026reinforcementlearningadaptivecomposition,
      title={Reinforcement Learning for Adaptive Composition of Quantum Circuit Optimisation Passes}, 
      author={Daniel Mills and Ifan Williams and Jacob Swain and Gabriel Matos and Enrico Rinaldi and Alexander Koziell-Pipe},
      year={2026},
      eprint={2601.21629},
      archivePrefix={arXiv},
      primaryClass={quant-ph},
      url={https://arxiv.org/abs/2601.21629}, 
}

@inproceedings{SC_similarity,
author = {Li, Lingda and Flynn, Thomas and Hoisie, Adolfy},
title = {Learning Generalizable Program and Architecture Representations for Performance Modeling},
year = {2024},
isbn = {9798350352917},
publisher = {IEEE Press},
url = {https://doi.org/10.1109/SC41406.2024.00072},
doi = {10.1109/SC41406.2024.00072},
booktitle = {Proceedings of the International Conference for High Performance Computing, Networking, Storage, and Analysis},
articleno = {66},
numpages = {15},
location = {Atlanta, GA, USA},
series = {SC '24}
}

@inproceedings{ICS_Similarity,
author = {Tr\"{u}mper, Lukas and Ben-Nun, Tal and Schaad, Philipp and Calotoiu, Alexandru and Hoefler, Torsten},
title = {Performance Embeddings: A Similarity-Based Transfer Tuning Approach to Performance Optimization},
year = {2023},
isbn = {9798400700569},
publisher = {Association for Computing Machinery},
address = {New York, NY, USA},
url = {https://doi.org/10.1145/3577193.3593714},
doi = {10.1145/3577193.3593714},
booktitle = {Proceedings of the 37th ACM International Conference on Supercomputing},
pages = {50–62},
numpages = {13},
location = {Orlando, FL, USA},
series = {ICS '23}
}

@misc{qiu2026passbypassoptimizationintentdrivenir,
      title={Beyond Pass-by-Pass Optimization: Intent-Driven IR Optimization with Large Language Models}, 
      author={Lei Qiu and Zi Yang and Fang Lyu and Ming Zhong and Huimin Cui and Xiaobing Feng},
      year={2026},
      eprint={2602.18511},
      archivePrefix={arXiv},
      primaryClass={cs.PL},
      url={https://arxiv.org/abs/2602.18511}, 
}

@inproceedings{quantum_tuning_cibda,
author = {Liu, Yi and Jin, Yuqiong and Xu, Jinchen},
title = {A Portable Auto-Tuning Framework for Quantum Compilation Optimization Based on D3QN},
year = {2025},
isbn = {9798400713163},
publisher = {Association for Computing Machinery},
address = {New York, NY, USA},
url = {https://doi.org/10.1145/3746709.3746950},
doi = {10.1145/3746709.3746950},
booktitle = {Proceedings of the 2025 6th International Conference on Computer Information and Big Data Applications},
pages = {1422–1428},
numpages = {7},
location = {
},
series = {CIBDA '25}
}

@article{osti_1623945,
title = {{A Variational Eigenvalue Solver on a Photonic Quantum Processor}},
author = {Peruzzo, Alberto and McClean, Jarrod and Shadbolt, Peter and Yung, Man-Hong and Zhou, Xiao-Qi and Love, Peter J. and Aspuru-Guzik, Alán and O’Brien, Jeremy L.},
abstractNote = {Quantum computers promise to efficiently solve important problems that are intractable on a conventional computer. For quantum systems, where the physical dimension grows exponentially, finding the eigenvalues of certain operators is one such intractable problem and remains a fundamental challenge. The quantum phase estimation algorithm efficiently finds the eigenvalue of a given eigenvector but requires fully coherent evolution. Here we present an alternative approach that greatly reduces the requirements for coherent evolution and combine this method with a new approach to state preparation based on ansa¨tze and classical optimization. We implement the algorithm by combining a highly reconfigurable photonic quantum processor with a conventional computer. We experimentally demonstrate the feasibility of this approach with an example from quantum chemistry—calculating the ground-state molecular energy for He–H þ . The proposed approach drastically reduces the coherence time requirements, enhancing the potential of quantum resources available today and in the near future.},
doi = {10.1038/ncomms5213},
journal = {Nature Communications},
number = {1},
volume = {5},
place = {United States},
year = {2014},
numpages={7},
month = {7}
}

@article{tosem_circuits,
author = {Ma, Ning and Li, Heng},
title = {Understanding and Estimating the Execution Time of Quantum Circuits},
year = {2025},
publisher = {Association for Computing Machinery},
address = {New York, NY, USA},
issn = {1049-331X},
url = {https://doi.org/10.1145/3778031},
doi = {10.1145/3778031},
note = {Just Accepted},
journal = {ACM Trans. Softw. Eng. Methodol.},
month = nov
}

@article{PhysRevLett_qa,
  title = {{Quantum Algorithm for Linear Systems of Equations}},
  author = {Harrow, Aram W. and Hassidim, Avinatan and Lloyd, Seth},
  journal = {Phys. Rev. Lett.},
  volume = {103},
  issue = {15},
  pages = {150502},
  numpages = {4},
  year = {2009},
  month = {Oct},
  publisher = {American Physical Society},
  doi = {10.1103/PhysRevLett.103.150502},
  url = {https://link.aps.org/doi/10.1103/PhysRevLett.103.150502}
}

@misc{openqasm2,
      title={Open Quantum Assembly Language}, 
      author={Andrew W. Cross and Lev S. Bishop and John A. Smolin and Jay M. Gambetta},
      year={2017},
      eprint={1707.03429},
      archivePrefix={arXiv},
      primaryClass={quant-ph},
      url={https://arxiv.org/abs/1707.03429}, 
}

@misc{chen2022veriqbenchbenchmarkmultipletypes,
      title={VeriQBench: A Benchmark for Multiple Types of Quantum Circuits}, 
      author={Kean Chen and Wang Fang and Ji Guan and Xin Hong and Mingyu Huang and Junyi Liu and Qisheng Wang and Mingsheng Ying},
      year={2022},
      eprint={2206.10880},
      archivePrefix={arXiv},
      primaryClass={quant-ph},
      url={https://arxiv.org/abs/2206.10880}, 
}

@INPROCEEDINGS{autophase,
  author={Huang, Qijing and Haj-Ali, Ameer and Moses, William and Xiang, John and Stoica, Ion and Asanovic, Krste and Wawrzynek, John},
  booktitle={2019 IEEE 27th Annual International Symposium on Field-Programmable Custom Computing Machines (FCCM)}, 
  title={AutoPhase: Compiler Phase-Ordering for HLS with Deep Reinforcement Learning}, 
  year={2019},
  volume={},
  number={},
  pages={308-308},
  doi={10.1109/FCCM.2019.00049}}

@inproceedings{oopsla_phase,
author = {Kulkarni, Sameer and Cavazos, John},
title = {Mitigating the compiler optimization phase-ordering problem using machine learning},
year = {2012},
isbn = {9781450315616},
publisher = {Association for Computing Machinery},
address = {New York, NY, USA},
url = {https://doi.org/10.1145/2384616.2384628},
doi = {10.1145/2384616.2384628},
booktitle = {Proceedings of the ACM International Conference on Object Oriented Programming Systems Languages and Applications},
pages = {147–162},
numpages = {16},
location = {Tucson, Arizona, USA},
series = {OOPSLA '12}
}

@article{QASMBench,
author = {Li, Ang and Stein, Samuel and Krishnamoorthy, Sriram and Ang, James},
title = {QASMBench: A Low-Level Quantum Benchmark Suite for NISQ Evaluation and Simulation},
year = {2023},
issue_date = {June 2023},
publisher = {Association for Computing Machinery},
address = {New York, NY, USA},
volume = {4},
number = {2},
url = {https://doi.org/10.1145/3550488},
doi = {10.1145/3550488},
journal = {ACM Transactions on Quantum Computing},
month = feb,
articleno = {10},
numpages = {26}
}

@misc{Web:qiskit,
	author = {{IBM}},
    title  = {{Introduction to Qiskit and IBM Quantum}},
	year   = {2026},
    note   = {Accessed: Apr 15th 2026. [Online]. Available: \url{https://quantum.cloud.ibm.com/docs/en/guides/}}
}

@misc{Web:gemini,
	author = {{Google DeepMind}},
    title  = {{Gemini 3.5 Flash}},
	year   = {2026},
    note   = {Accessed: Jun 5th 2026. [Online]. Available: \url{https://deepmind.google/models/model-cards/gemini-3-5-flash/}}
}

@misc{Web:dsv32,
    author = {{DeepSeek}},
    title  = {{DeepSeek-V3.2 Release}},
    year   = {2025},
    note   = {Accessed: Jun 5th 2026. [Online]. Available: \url{https://api-docs.deepseek.com/news/news251201/}}
}

@misc{Web:dsv4,
	author = {{DeepSeek}},
    title  = {{DeepSeek V4 Preview Release}},
	year   = {2026},
    note   = {Accessed: Jun 5th 2026. [Online]. Available: \url{https://api-docs.deepseek.com/news/news260424/}}
}

@misc{Web:gemini31,
	author = {{Google DeepMind}},
    title  = {{Gemini 3.1 Flash-Lite Model Card}},
	year   = {2026},
    note   = {Accessed: Jun 5th 2026. [Online]. Available: \url{https://deepmind.google/models/model-cards/gemini-3-1-flash-lite/}}
}

@misc{Web:doubao,
    author = {{ByteDance Seed}},
    title  = {{Seed 2.0}},
    year   = {2026},
    note   = {Accessed: Jun 24th 2026. [Online]. Available: \url{https://seed.bytedance.com/en/seed2}}
}

@misc{Web:glm5,
    author = {{Zhipu AI}},
    title  = {{GLM-5: Next Generation Large Language Model}},
    year   = {2026},
    note   = {Accessed: Jun 24th 2026. [Online]. Available: \url{https://glm-5.org/}}
}

@misc{Web:claude,
    author = {{Claude}},
    title  = {{Introducing Claude Opus 5}},
    year   = {2026},
    note   = {Accessed: Jun 25th 2026. [Online]. Available: \url{https://www.anthropic.com/news/claude-opus-5}}
}

@misc{Web:gpt55,
  author = {{OpenAI}},
  title  = {{Introducing GPT-5.5}},
  year   = {2026},
  note   = {Accessed: Jun 4th 2026. [Online]. Available: \url{https://openai.com/introducing-gpt-5-5/}}
}

@misc{Web:gpt5,
  author = {{OpenAI}},
  title  = {{GPT-5 is here}},
  year   = {2025},
  note   = {Accessed: Jun 2nd 2026. [Online]. Available: \url{https://openai.com/gpt-5/}}
}

@misc{Web:gpt52,
  author = {{OpenAI}},
  title  = {{Introducing GPT-5.2}},
  year   = {2025},
  note   = {Accessed: Jul 1st 2026. [Online]. Available: \url{https://openai.com/introducing-gpt-5-2}}
}

@misc{Web:qwen35,
  author = {{QWen}},
  title  = {{Qwen3.5: Towards Native Multimodal Agents}},
  year   = {2026},
  note   = {Accessed: Jun 5th 2026. [Online]. Available: \url{https://qwen.ai/blog?id=qwen3.5}}
}

@misc{Web:openrouter,
  author = {{OpenRouter}},
  title  = {{The Unified Interface For Every Model}},
  year   = {2026},
  note   = {Accessed: May 4th 2026. [Online]. Available: \url{https://openrouter.ai}}
}

@misc{Web:claude46,
  author = {{Anthropic}},
  title  = {{Introducing Claude Sonnet 4.6}},
  year   = {2026},
  note   = {Accessed: Jun 8th 2026. [Online]. Available: \url{https://www.anthropic.com/news/claude-sonnet-4-6}}
}

@inbook{significance_test,
author = {Woolson, R. F.},
publisher = {John Wiley \& Sons, Ltd},
isbn = {9780471462422},
title = {Wilcoxon Signed-Rank Test},
booktitle = {Wiley Encyclopedia of Clinical Trials},
chapter = {},
pages = {1-3},
year = {2008},
}

@misc{fidelity,
      title={{Quantum Computing with Qiskit}}, 
      author={Ali Javadi-Abhari and Matthew Treinish and Kevin Krsulich and Christopher J. Wood and Jake Lishman and Julien Gacon and Simon Martiel and Paul D. Nation and Lev S. Bishop and Andrew W. Cross and Blake R. Johnson and Jay M. Gambetta},
      year={2024},
      eprint={2405.08810},
      archivePrefix={arXiv},
      primaryClass={quant-ph},
      url={https://arxiv.org/abs/2405.08810}, 
}

@article{ertree,
author = {Geurts, Pierre and Ernst, Damien and Wehenkel, Louis},
title = {Extremely randomized trees},
year = {2006},
issue_date = {April 2006},
publisher = {Kluwer Academic Publishers},
address = {USA},
volume = {63},
number = {1},
issn = {0885-6125},
url = {https://doi.org/10.1007/s10994-006-6226-1},
doi = {10.1007/s10994-006-6226-1},
journal = {Mach. Learn.},
month = apr,
pages = {3–42},
numpages = {40}
}

@article{zhong2026qutunerfeaturelearningguidedoptimization,
      title={QuTuner: Feature- and Learning-Guided Optimization Pass Tuning for Quantum Compilers}, 
      author={Ming Zhong and Xiangyu Ren and Jinglei Cheng and Shaohua Li and Zhiding Liang},
      year={2026},
      journal = {arXiv preprint arXiv:2607.04586},
      primaryClass={quant-ph},
      url={https://arxiv.org/abs/2607.04586}, 
}

@article{quantum_chem,
author = {Cao, Yudong and Romero, Jonathan and Olson, Jonathan P. and Degroote, Matthias and Johnson, Peter D. and Kieferová, Mária and Kivlichan, Ian D. and Menke, Tim and Peropadre, Borja and Sawaya, Nicolas P. D. and Sim, Sukin and Veis, Libor and Aspuru-Guzik, Alán},
title = {{Quantum Chemistry in the Age of Quantum Computing}},
journal = {Chemical Reviews},
volume = {119},
number = {19},
pages = {10856-10915},
year = {2019},
doi = {10.1021/acs.chemrev.8b00803},
note ={PMID: 31469277},
URL = {https://doi.org/10.1021/acs.chemrev.8b00803},
eprint = { https://doi.org/10.1021/acs.chemrev.8b00803}
}

@inproceedings{quest,
author = {Wang, Hanrui and Liang, Zhiding and Gu, Jiaqi and Li, Zirui and Ding, Yongshan and Jiang, Weiwen and Shi, Yiyu and Pan, David Z. and Chong, Frederic T. and Han, Song},
title = {TorchQuantum Case Study for Robust Quantum Circuits},
year = {2022},
isbn = {9781450392174},
publisher = {Association for Computing Machinery},
address = {New York, NY, USA},
url = {https://doi.org/10.1145/3508352.3561118},
doi = {10.1145/3508352.3561118},
booktitle = {Proceedings of the 41st IEEE/ACM International Conference on Computer-Aided Design},
articleno = {136},
numpages = {9},
location = {San Diego, California},
series = {ICCAD '22}
}

@article{qse_roadmap,
author = {Murillo, Juan Manuel and Garcia-Alonso, Jose and Moguel, Enrique and Barzen, Johanna and Leymann, Frank and Ali, Shaukat and Yue, Tao and Arcaini, Paolo and P\'{e}rez-Castillo, Ricardo and Garc\'{\i}a-Rodr\'{\i}guez de Guzm\'{a}n, Ignacio and Piattini, Mario and Ruiz-Cort\'{e}s, Antonio and Brogi, Antonio and Zhao, Jianjun and Miranskyy, Andriy and Wimmer, Manuel},
title = {Quantum Software Engineering: Roadmap and Challenges Ahead},
year = {2025},
issue_date = {June 2025},
publisher = {Association for Computing Machinery},
address = {New York, NY, USA},
volume = {34},
number = {5},
issn = {1049-331X},
url = {https://doi.org/10.1145/3712002},
doi = {10.1145/3712002},
journal = {ACM Trans. Softw. Eng. Methodol.},
month = may,
articleno = {154},
numpages = {48}
}

@ARTICLE{qse_test1,
  author={Wang, Xinyi and Ali, Shaukat and Yue, Tao and Arcaini, Paolo},
  journal={IEEE Transactions on Software Engineering}, 
  title={Quantum Approximate Optimization Algorithm for Test Case Optimization}, 
  year={2024},
  volume={50},
  number={12},
  pages={3249-3264},
  doi={10.1109/TSE.2024.3479421}
}

@article{qse_test2,
author = {Xia, Shangzhou and Zhao, Jianjun and Zhang, Fuyuan and Guo, Xiaoyu},
title = {Quantum Concolic Testing},
year = {2025},
issue_date = {July 2025},
publisher = {Association for Computing Machinery},
address = {New York, NY, USA},
volume = {2},
number = {ISSTA},
url = {https://doi.org/10.1145/3728926},
doi = {10.1145/3728926},
journal = {Proc. ACM Softw. Eng.},
month = jun,
articleno = {ISSTA051},
numpages = {21}
}

@article{qse_test3,
author = {Long, Peixun and Zhao, Jianjun},
title = {Testing Multi-Subroutine Quantum Programs: From Unit Testing to Integration Testing},
year = {2024},
issue_date = {July 2024},
publisher = {Association for Computing Machinery},
address = {New York, NY, USA},
volume = {33},
number = {6},
issn = {1049-331X},
url = {https://doi.org/10.1145/3656339},
doi = {10.1145/3656339},
journal = {ACM Trans. Softw. Eng. Methodol.},
month = jun,
articleno = {147},
numpages = {61}
}

@article{qse_repair,
author = {Tan, Siwei and Lu, Liqiang and Xiang, Debin and Chu, Tianyao and Lang, Congliang and Chen, Jintao and Hu, Xing and Yin, Jianwei},
title = {HornBro: Homotopy-Like Method for Automated Quantum Program Repair},
year = {2025},
issue_date = {July 2025},
publisher = {Association for Computing Machinery},
address = {New York, NY, USA},
volume = {2},
number = {FSE},
url = {https://doi.org/10.1145/3715751},
doi = {10.1145/3715751},
journal = {Proc. ACM Softw. Eng.},
month = jun,
articleno = {FSE034},
numpages = {23}
}

@inproceedings{tuniq,
author = {Mohammad Abrarul Hasanat and Jason, Ludmir and Tirthak, Patel and Rohan Basu Roy},
title = {TuniQ: Autotuning Compilation Passes for Quantum Workloads at Scale for Effectiveness and Efficiency},
year = {2026},
publisher = {Association for Computing Machinery},
address = {New York, NY, USA},
booktitle = {Proceedings of the 39th ACM International Conference on Supercomputing},
location = {Belfast, Northern Ireland},
series = {ICS '26}
}

@INPROCEEDINGS{ASE_Quanbench,
  author={Xiaoyu Guo and Minggu Wang and Jianjun Zhao},
  booktitle={2025 40th IEEE/ACM International Conference on Automated Software Engineering (ASE)}, 
  title={QuanBench: Benchmarking Quantum Code Generation with Large Language Models}, 
  year={2025},
  publisher = {IEEE Press},
  address = {Seoul, South Korea},
  volume={},
  number={},
}

@INPROCEEDINGS{qse_test4,
  author={Wang, Jiyuan and Zhang, Qian and Xu, Guoqing Harry and Kim, Miryung},
  booktitle={2021 36th IEEE/ACM International Conference on Automated Software Engineering (ASE)}, 
  title={QDiff: Differential Testing of Quantum Software Stacks}, 
  year={2021},
  volume={},
  number={},
  publisher = {IEEE Press},
  address = {Melbourne, Australia},
  pages={692-704},
  doi={10.1109/ASE51524.2021.9678792}
}

@inproceedings{qse_codegen1,
   title={PennyCoder: Efficient Domain-Specific LLMs for PennyLane-Based Quantum Code Generation},
   url={http://dx.doi.org/10.1109/QCE65121.2025.10324},
   DOI={10.1109/qce65121.2025.10324},
   booktitle={2025 IEEE International Conference on Quantum Computing and Engineering (QCE)},
   publisher={IEEE},
   address = {Albuquerque, New Mexico, USA},
   author={Basit, Abdul and Shao, Minghao and Asif, Muhammad Haider and Innan, Nouhaila and Kashif, Muhammad and Marchisio, Alberto and Shafique, Muhammad},
   year={2025},
   month=aug, pages={229–234}
}

@inproceedings{qse_codegen2,
author = {Campbell, Charlie and Chen, Hao (Mark) and Luk, Wayne and Fan, Hongxiang},
title = {Enhancing LLM-Based Quantum Code Generation with Multi-Agent Optimization and Quantum Error Correction},
year = {2025},
isbn = {9798331503048},
publisher = {IEEE Press},
url = {https://doi.org/10.1109/DAC63849.2025.11133316},
doi = {10.1109/DAC63849.2025.11133316},
booktitle = {Proceedings of the 62nd Annual ACM/IEEE Design Automation Conference},
articleno = {414},
numpages = {7},
address = {San Francisco, California, United States},
series = {DAC '25}
}

@inproceedings{liang2024combining,
  title={Combining Parameterized Pulses and Contextual Subspace for More Practical VQE},
  author={Liang, Zhiding and Song, Zhixin and Cheng, Jinglei and Ren, Hang and Hao, Tianyi and Yang, Rui and Shi, Yiyu and Li, Tongyang},
  booktitle={Proceedings of the 61st ACM/IEEE Design Automation Conference},
  pages={1--6},
  year={2024}
}
\end{document}